\documentclass[lettersize,journal]{IEEEtran}
\usepackage{amsmath,amsfonts}
\usepackage{array}
\usepackage[caption=false,font=normalsize,labelfont=sf,textfont=sf]{subfig}
\usepackage{textcomp}
\usepackage{stfloats}
\usepackage{url}
\usepackage{verbatim}
\usepackage{graphicx}
\usepackage{cite}
\usepackage{enumitem}

\usepackage{color}
\usepackage{tikz}
\usepackage{amsmath}
\usepackage[ruled,linesnumbered]{algorithm2e}
\usepackage{flexisym}
\usepackage{breqn}
\usepackage{hyperref}
\usepackage{cleveref}
\usepackage{colortbl}
\usepackage{booktabs}
\usepackage[skins]{tcolorbox}

\usepackage{xspace}
\usepackage{pifont}
\usepackage{caption}
\usepackage{makecell}
\usepackage{threeparttable}
\usepackage{multirow}
\usepackage{multicol}

\usepackage{xcolor}
\usepackage{float} 
\usepackage{fontawesome}
\usepackage{tabularx}
\usepackage[caption=false]{subfig}
\usepackage[normalem]{ulem}

\usepackage[adversary,operators,sets,events, keys, probability]{cryptocode}

\crefformat{section}{\S#2#1#3} 
\crefformat{subsection}{\S#2#1#3}
\crefformat{subsubsection}{\S#2#1#3}

\renewcommand{\arraystretch}{1.2}

\newtcolorbox{mybox}[2][]{colback=white, colframe=black, coltitle=black, fonttitle=\bfseries, boxrule=1.0pt,
colbacktitle=white, enhanced, attach boxed title to top right={xshift=-5mm, yshift=-2mm}, after skip=0pt, title={#2},#1}

\definecolor{bred}{rgb}{0.8, 0.0, 0.0}
\definecolor{cgreen}{rgb}{0.0, 0.42, 0.24}

\newcommand\colourcheck{{\textcolor{cgreen}{\ding{51}}}}
\newcommand\colourxmark{{\textcolor{bred}{\ding{55}}}}
\newcommand\tbNA{\textcolor{gray}{N/A}}

\newcommand{\vrf}{\mathcal{V}\mathit{rf}}
\newcommand{\att}{\mathcal{A}\mathit{tt}}
\newcommand{\manu}{\mathcal{M}\mathit{fr}}
\newcommand{\srpr}{\mathcal{R}\mathcal{P}}
\newcommand{\adver}{\mathcal{A}\mathit{dv}}

\newcommand{\id}{\mathsf{id}}
\newcommand{\aid}{\mathsf{aid}}
\newcommand{\vid}{\mathsf{vid}}
\newcommand{\gid}{\mathsf{gid}}
\newcommand{\pid}{\mathsf{pid}}
\newcommand{\ssid}{\mathsf{ssid}}
\newcommand{\prog}{\mathsf{prog}}
\newcommand{\Ss}{\mathcal{S}}
\newcommand{\Aa}{\mathcal{A}}

\newcommand{\Ii}{\mathcal{I}}
\newcommand{\Kk}{\mathcal{K}}
\newcommand{\Ww}{\mathcal{W}}
\newcommand{\Ll}{\mathcal{L}}
\newcommand{\kid}{\mathcal{K}_{\ i}}
\newcommand{\iid}{\mathcal{I}_{i}}
\newcommand{\wid}{\mathcal{W}_{i}}
\newcommand{\janus}{{\scshape Janus}}
\newcommand{\puf}{\textit{puf}}
\newcommand{\mli}[1]{\mathit{#1}}

\newtheorem{theorem}{Theorem}

\newtheorem{definition}{Definition}
\newenvironment{proof}{{\noindent\it Proof}\quad}{\hfill $\square$\par}

\makeatletter
\DeclareRobustCommand\onedot{\futurelet\@let@token\@onedot}
\def\@onedot{\ifx\@let@token.\else.\null\fi\xspace}

\def\eg{\emph{e.g}\onedot} 

\def\ie{\emph{i.e}\onedot}

\def\etal{\emph{et al}\onedot}

\definecolor{cardinal}{rgb}{0.77, 0.12, 0.23}
\definecolor{ceruleanblue}{rgb}{0.16, 0.32, 0.75}
\definecolor{lavenderblue}{rgb}{0.8, 0.8, 1.0}
\definecolor{janusblue}{RGB}{96, 136, 172}
\definecolor{janusbrown}{RGB}{132, 78, 1}

\newcommand*{\priority}[1]{\begin{tikzpicture}[scale=0.13]%
		\draw (0,0) circle (1);
		\fill[fill opacity=1,fill=janusblue] (0,0) -- (90:1) arc (90:90-#1*3.6:1) -- cycle;
\end{tikzpicture}}

\newcommand{\nosemic}{\renewcommand{\@endalgocfline}{\relax}}
\let\oldnl\nl
\newcommand{\nonl}{\renewcommand{\nl}{\let\nl\oldnl}}

\SetCommentSty{mycommfont}
\SetArgSty{textnormal}
\SetKwInOut{Input}{Input}
\SetKwInOut{Output}{Output}
\SetKw{False}{false}
\SetKw{True}{true}
\SetKw{Valid}{valid}
\SetKw{Invalid}{invalid}

\newcommand{\compfull}{\priority{100}}
\newcommand{\comppart}{\priority{50}}
\newcommand{\compnone}{\priority{0}}

\begin{document}

\title{Teamwork Makes TEE Work: Open and Resilient Remote Attestation on Decentralized Trust}

\author{Xiaolin Zhang, Kailun Qin, Shipei Qu, Tengfei Wang, Chi Zhang, Dawu Gu
\thanks{Xiaolin Zhang, Kailun Qin, Shipei Qu, Tengfei Wang, Chi Zhang and Dawu Gu are with School of Electronic Information and Electrical Engineering, Shanghai Jiao Tong University (SJTU), Shanghai 200240, China (email: xiaolinzhang@sjtu.edu.cn; kailun.qin@sjtu.edu.cn; shipeiqu@sjtu.edu.cn; tengfei2019@sjtu.edu.cn; zcsjtu@sjtu.edu.cn; dwgu@sjtu.edu.cn).}
\thanks{Corresponding Author: Chi Zhang, Dawu Gu}
\thanks{Manuscript received XXX, XXX; revised XXX, XXX.}}

\markboth{Journal of \LaTeX\ Class Files,~Vol.~14, No.~8, August~2021}%
{Shell \MakeLowercase{\textit{et al.}}: A Sample Article Using IEEEtran.cls for IEEE Journals}

\IEEEpubid{0000--0000/00\$00.00~\copyright~2021 IEEE}

\maketitle

\begin{abstract}
  Remote Attestation (RA) enables the integrity and authenticity of applications in Trusted Execution Environment (TEE) to be verified. Existing TEE RA designs employ a centralized trust model where they rely on a single provisioned secret key and a centralized verifier to establish trust for remote parties. This model is however brittle and can be untrusted under advanced attacks nowadays. Besides, most designs only have fixed procedures once deployed, making them hard to adapt to different emerging situations and provide resilient functionalities. 

  Therefore, we propose \janus, an open and resilient TEE RA scheme. To decentralize trust, we, on one hand, introduce Physically Unclonable Function (PUF) as an intrinsic root of trust (RoT) in TEE to directly provide physical trusted measurements. On the other hand, we design novel decentralized verification functions on smart contract with result audits and RA session snapshot. Furthermore, we design an automated switch mechanism that allows \janus\ to remain resilient and offer flexible RA services under various situations. We provide a UC-based security proof and demonstrate the scalability and generality of \janus\ by implementing an complete prototype.
\end{abstract}

\begin{IEEEkeywords}
Trusted execution environment, Remote attestation, Physically unclonable function, Blockchain, Smart contract.
\end{IEEEkeywords}

\section{Introduction}\label{sec:introduction}

\IEEEPARstart{T}{rusted} Execution Environment (TEE) has emerged as a cornerstone of protecting sensitive applications by offering isolated areas (\eg, \textit{enclaves}). Remote Attestation (RA) is an essential component to ensure the integrity and authenticity of TEEs, particularly in scenarios where trust must be established between remote parties. In the standard RA workflow \cite{RATSrfc}, a verifier $\vrf$ checks whether the application (the attester $\att$) in a remote TEE device is authentic. $\vrf$ verifies $\att$'s measurement, which is signed by an attestation key derived from a provisioned root key. Through RA, various stakeholders, such as cloud service providers, data owners, and developers, can use TEEs assuredly in hostile environments.

RA is necessary for modern TEE platforms. It helps establish trust relationships and secure channels, serving as a preliminary step before launching actual computing tasks within the TEE. Therefore, nearly all cloud vendors and modern TEEs, such as Azure \cite{azurera}, AWS \cite{awsra}, and SGX \cite{costan2016intel}, Sanctum \cite{costan2016sanctum}, Keystone \cite{leeKeystoneOpenFramework2020}, provide RA functions. However, we notice that current designs exhibit three significant limitations in practice.

$\bullet$\ \textbf{Untrusted TEE Platforms}: Current TEE RA designs rely on a \textbf{centralized} and \textbf{manufacturer-dominated} root of trust (RoT), \ie, the hardware root key provisioned by the TEE manufacturer. Users have to trust the manufacturers to trust the keys. However, in recent years, side-channel attacks \cite{kocher2020spectre,Lipp0G0HFHMKGYH18Meltdown, schaik2021cacheout,PLATYPUSLippKOSECG21}, fault injection attacks \cite{CLKSCREWTangSS17,BADFETCuiH17, VoltJockeyQiuWLQ19,GlitchBuhrenJKS21}, and other real-world exploits \cite{sgxfail,LVISPAttestation,ZombieLoad} have proven that protecting this centralized and digitally provisioned RoT is quite challenging against advanced attackers. Once the key is leaked, attackers can generate fake RA reports at will \cite{sgaxe} to cover up that the TEE platform has been completely compromised.

$\bullet$\ \textbf{Untrusted Verifier}: The verification procedure in RA is conducted by a centralized verifier \cite{costan2016intel, azurera, awsra, googlera}, lacking \textbf{transparency} and \textbf{open inspection}. Users cannot verify or audit the genuineness of the verification results. The centralized verifier could claim a compromised application to be trusted or vice versa out of business interests or external attacks. It is even worse that the verification services are deployed by the manufacturer itself \cite{intelIAS}. The monolithic trust model can sabotages users' trust on TEEs \cite{sgaxe,chenOPERAOpenRemote2019}.

This \textit{centralized trust} on both the attester and verifier sides can cause serious vulnerabilities to the entire RA procedure, \eg, malicious application measurements, fake signed reports, and unverifiable results. One part being compromised in the chain of trust lead to the failure of the whole system. This fundamental problem prevents RA from fulfilling its very purpose, \ie, establishing authentic trust between remote entities.

$\bullet$\ \textbf{Rigid Procedure}: Most TEE RA designs follow a \textbf{fixed} procedure \cite{RATSrfc} for standard scenarios and depend heavily on the underlying cloud infrastructure to function properly \cite{azurera,awsra}. They lack the resilience \cite{NISTResilience} to remain operational under adverse conditions, which has been required in various systems \cite{mscloudnative, msazureresiliency, googleinfrastructure, awsarchitecture}. Such resilience is crucial since RA must be performed before most other TEE functions. Real-world incidents such as network outages, device malfunctions and computing exhaustion can occur, necessitating RA to provide resilient functionalities to continuously ensure the platforms integrity. As TEEs are increasingly used in diverse emerging scenarios, there is a stronger incentive for RA designs to address this aspect and avoid using a rigid workflow.

\IEEEpubidadjcol

According to these limitations, we identify three crucial requirements for designing an open and resilient TEE RA scheme with end-to-end trust and high-available attestation functionality. First \textbf{(R1)}, the growing number of advanced attacks \cite{lipp2016armageddon, kocher2020spectre,Lipp0G0HFHMKGYH18Meltdown, schaik2021cacheout,PLATYPUSLippKOSECG21, CLKSCREWTangSS17,BADFETCuiH17, VoltJockeyQiuWLQ19,GlitchBuhrenJKS21, sgaxe,sgxfail} in recent years make it urgent for TEE RA to adopt a stronger RoT with independent trust for attestation, instead of relying on a centralized root key. Second \textbf{(R2)}, the vendor-centric single-point-of-verification model poses great risks of losing trust or being compromised by external attackers \cite{chenOPERAOpenRemote2019}. It is essential to create a more open procedure that supports various trusted verification functions and multi-party participation. Third \textbf{(R3)}, it is also necessary for users to select different RA workflows and additional functions to handle unexpected situations, ensuring that the attestation procedure can bootstrap trust for other tasks in TEE.

\noindent\textbf{Existing Approaches.} Unfortunately, despite tremendous efforts in TEE RA protocol designs, none of the existing works satisfy all three requirements at once. Table \ref{tab:design_compare} presents the representative designs in industry and academia.

Most designs rely on a root key that is embedded by the manufacturer or generated by hardware \cite{costan2016sanctum,RomanPUFHash}. The entire RA procedure can be immediately compromised once the key string is leaked by various attacks \cite{sgxfail,sgaxe, GlitchBuhrenJKS21, PLATYPUSLippKOSECG21,gofetch, ghostraceCVE,LVISPAttestation,schaik2021cacheout}, regardless of how it is created (not satisfying \textbf{R1}). On the verifier side, some designs \cite{chenOPERAOpenRemote2019, petziSCRAPSScalableCollective2022,ankergardPERMANENTPubliclyVerifiable2022} do allow more parties to perform the verification together. However, they only provide signature verification and lack rich functions to build a more open procedure. Such decentralized function is insufficient in emerging scenarios \cite{googlefaas,intelamber,googlebeyondcorp,CISAInternetConnections} (not fully satisfying \textbf{R2}). Meanwhile, almost no research \cite{galanouMATEEMultimodalAttestation2022} have explored how to provide the native resilience for a complete RA procedure. They only offer specific, fixed attestation workflows, requiring a stable cloud environment for users to verify their applications in TEE (not satisfying \textbf{R3}).

\begin{table*}[t]
    \footnotesize
	\centering
	\caption{Comparison between \janus\ and related works \label{tab:design_compare}}
    \begin{threeparttable}
	\begin{tabular}{ccccccc}
		\toprule
        \multirow{2}{*}{\textbf{Design}} & \multicolumn{2}{c}{\textbf{R1: Trusted RA RoT}} & \multicolumn{2}{c}{\textbf{R2: Open Verification}} & \multicolumn{2}{c}{\textbf{R3: Resilient Functions}} \\
        \cmidrule(r){2-3}\cmidrule(r){4-5}\cmidrule(r){6-7}
        ~ & RoT Type & Intrinsic Trust$^{1}$ & Multi-party Verification & Auditable Results & Redundant Workflows & Infra. Independent \\
		\midrule
        \multicolumn{7}{c}{\cellcolor{gray!15}\textsc{Industry}}\\
        Intel IAS \cite{intelIAS} & \textcolor{bred}{root key} & \colourxmark & \colourxmark & \colourxmark & \colourxmark & \colourxmark \\
        Intel DCAP \cite{scarlataSupportingThirdParty} & \textcolor{bred}{root key} & \colourxmark & \colourcheck & \colourxmark & \colourxmark & \colourxmark \\
        Azure RA \cite{azurera} & \textcolor{bred}{root key} & \colourxmark & \colourxmark & \colourxmark & \colourxmark & \colourxmark \\
        Google RA \cite{googlera} & \textcolor{bred}{root key} & \colourxmark & \colourxmark & \colourxmark & \colourxmark & \colourxmark \\
        \hline
        \multicolumn{7}{c}{\cellcolor{gray!15}\textsc{Academia}}\\
        OPERA \cite{chenOPERAOpenRemote2019} & \textcolor{bred}{root key} & \colourxmark & \colourcheck & \colourxmark & \colourxmark & \colourcheck$^{3}$ \\
        Sanctum RA \cite{costan2016sanctum} & \textcolor{gray!75}{root key} & \colourcheck & \colourxmark & \colourxmark & \colourxmark & \colourxmark \\
        LIRA-V \cite{shepherdLIRAVLightweightRemote2022} & \textcolor{bred}{root key} & \colourxmark & \colourxmark & \colourxmark & \colourxmark & \colourxmark \\
        MATEE \cite{galanouMATEEMultimodalAttestation2022} & \textcolor{bred}{root key} & \colourxmark & \colourxmark & \colourxmark & \colourcheck$^{2}$ & \colourxmark \\
        SCRAPS \cite{petziSCRAPSScalableCollective2022} & \textcolor{bred}{root key} & \colourxmark & \colourcheck & \colourxmark & \colourxmark & \colourcheck \\
        PERMANENT \cite{ankergardPERMANENTPubliclyVerifiable2022} & \textcolor{bred}{root key} & \colourxmark & \colourcheck & \colourxmark & \colourxmark & \colourcheck \\
        SGXOnerated \cite{SGXonerated} & \textcolor{bred}{root key} & \colourxmark & \colourcheck & \colourxmark & \tbNA & \tbNA \\
        DAA \cite{brickell2004direct} & \textcolor{bred}{root key} & \colourxmark & \colourcheck & \colourxmark & \tbNA & \tbNA \\
        Rom{\'{a}}n \etal \cite{RomanPUFHash} & \textcolor{gray!75}{root key} & \colourcheck & \colourxmark & \colourxmark & \colourxmark & \colourxmark \\
        \textbf{\janus\ (this paper)} & \textcolor{cgreen}{PUF} & \colourcheck & \colourcheck & \colourcheck & \colourcheck & \colourcheck \\
		\bottomrule
	\end{tabular}
    \begin{tablenotes}
        \small 
        \setlength{\columnsep}{0.8cm}
        \setlength{\multicolsep}{0cm}
        \begin{multicols}{2}
            \item[{\makebox[0.8cm][r]{{\textcolor{bred}{root key}}}}]Relying on a manufacturer-provisioned device root key.
            \item[{\makebox[0.8cm][r]{\textcolor{gray!75}{root key}}}]Using PUF to generate the root key of TEE.
            \item[{\makebox[0.8cm][r]{\textcolor{cgreen}{PUF}}}]Directly embedding PUF into TEE RA functions.
            \item[1]Trust is not derived from any artificial entity. 
            \item[2]Having extra TPM-based RA functionality for SGX.
            \item[3]Third parties are allowed to deploy the verification services. 
        \end{multicols}
    \end{tablenotes}
    \end{threeparttable}
\end{table*}





\noindent\textbf{Our Design.} Although the three requirements may seem orthogonal to each other, we identify that the fundamental factors to meet them lie in \textit{decentralization} and \textit{redundancy}. Therefore, in this paper, we propose \janus\footnote{\janus\ is an ancient roman guardian god with two faces}, an open and resilient TEE RA design on decentralized end-to-end trust with flexible functions and procedures. Compared with other works, we fully integrate these two factors throughout the entire RA procedure of \janus, allowing it to satisfy the three crucial requirements simultaneously.

For \textbf{R1}, \janus\ decentralizes the centralized root key of TEE by Physically Unclonable Function (PUF) \cite{pappuPhysicalOneWayFunctions2002}. PUF is a hardware primitive that incorporates physical variations during manufacturing to its intrinsic structure, which cannot be determined even by its manufacturer. Some designs have introduced PUF into TEE/RA for key generation \cite{costan2016sanctum,RomanPUFHash} or application measurement \cite{ghaeiniPAttPhysicsbasedAttestation}. In \janus, we take a step further by offering this insight: PUF establishes \textbf{intrinsic trust} based on physics instead of human configuration. Itself genuinely decentralizes the centralized trust model on a single root key and address the untrusted platform limitation. \janus\ offers PUF-based mutual attestation protocols for local and remote enclaves in TEE. While the way PUF is used is somewhat similar with other research, we treat it as an additional stronger RoT with independent trust and the purpose is different. 


For \textbf{R2}, \janus\ decentralizes the centralized verification procedure by smart contracts. Indeed, many works in Table \ref{tab:design_compare} have adopted the same technique, \janus\ goes beyond them by offering \textit{attestation audit}, \textit{batch attestation} and other novel functions to enable a more flexible, auditable and lightweight procedure. They create a more open verification procedure in a stronger decentralized manner, thereby thoroughly addressing the untrusted verifier limitation. For \textbf{R3}, we design an automated switch mechanism using a dedicated smart contract in \janus. It allows users to select different workflows to conduct the attestation, \ie, using off-chain PUF-based protocols or on-chain RA functions. This two-fold architecture makes \janus\ continuously functional against incidents such as network outages, device malfunctions and computing exhaustion, ensuring flexible and high-reliable RA service delivery.


We evaluated \janus\ with a Proof-of-Concept (PoC) implementation. The system prototype includes an SGX-enabled Dell EMC server and an NXP LPC55S69 board with Hyperledger Sawtooth \cite{olsonSawtoothIntroduction2018}. The server is connected to an IPUF \cite{nguyen2019interpose}. The LPC board has a built-in SRAM PUF IP. They mutually attest with each other using the on/off-chain RA functions. Sawtooth is the backend of our verification service and switch mechanism. Our evaluation show that the off-chain attestation using PUF and TEE take $\sim$5 ms on EMC Server and $\sim$510 ms on LPC board (\textbf{R1}). The on-chain verification functions take 8.3 ms$\sim$30 ms per transaction (\textbf{R2}). For the attestation switch, \janus\ provides a policy-based approach which requires a 5.7KB contract and incurs an on-chain latency of 4.4 ms (\textbf{R3}).

\noindent\textbf{Contributions.} \janus\ is the first TEE RA design with end-to-end trust and resilient workflow on decentralized RoT and verification procedure. Specifically, our work contributes by:



\begin{itemize}[itemsep=0pt, parsep=0pt, topsep=2pt]
    \item We use the independent trust from PUF to decentralize the single trust provided by artificial root keys in TEE for a more trusted and stronger RoT for RA. We design PUF-based mutual attestation protocols for applications in local or remote enclaves. We provide a UC-based \cite{canetti2001universally} formal security proof for the protocols. 
    \item We design novel smart contract-based RA functions for transparent decentralized verification. They enable efficient on-chain batch attestation and the audits of historical results by snapshots. creating a more open RA verification procedure for practical applications.
    \item We propose a switch mechanism to combine the PUF-based protocols and smart contract attestation. It ensures automated workflow transitions between off-chain and on-chain RA functions in the event of unexpected system failures, thereby achieving resilience by design.
    \item We give a complete PoC implementation of \janus\footnote{It will be open-sourced at \url{https://sites.google.com/view/janus-ra}}. The comparison with other RA schemes show that \janus\ has small storage overhead and achieves great scalability and reasonable computing performance.
\end{itemize}


\noindent\textbf{Scope.} We focus on the system design of TEE RA schemes in this study to address the issues of over-centralization and lack of resilience, which exist in most designs and have grown as urgent industrial concerns. We assume that the off-the-shelf hardware environments with PUF-embedded TEEs are available. The hardware integration of PUF into existing TEEs is out of scope of this work. 

\section{Background}\label{sec:background}

\subsection{Trusted Execution Environment and Remote Attestation}\label{sec:tee_ra_background}


\textbf{Trusted Execution Environment (TEE)} aims to create a secure, isolated environment within a processor (xPU) where sensitive code can execute securely without interference from the normal OS. It typically requires hardware and OS-level support to ensure the integrity and confidentiality of the loaded applications. TEEs can be roughly classified into enclave-based (Intel SGX \cite{costan2016intel}, Sanctum \cite{costan2016sanctum}, Keystone \cite{leeKeystoneOpenFramework2020}, PENGLAI \cite{PENGLAITEE}), virtualization-based (ARM Trustzone, Intel TDX \cite{IntelTrustDomain}, AMD SEV \cite{amdsev}, ARM CCA \cite{armcca}).

In recent years, TEEs have encountered numerous advanced attacks \cite{lipp2016armageddon, kocher2020spectre,Lipp0G0HFHMKGYH18Meltdown, schaik2021cacheout,PLATYPUSLippKOSECG21, CLKSCREWTangSS17,BADFETCuiH17, VoltJockeyQiuWLQ19,GlitchBuhrenJKS21, sgxfail, sgaxe} in the real world. These attacks can expose the isolated secrets and compromise the applications of other tenants in cloud environments. Therefore, before executing any actual tasks in TEE, it is necessary to verify the integrity of the loaded program to ensure the security of TEE itself.

\begin{figure}[htbp]
  \centering
  \includegraphics[width=0.9\linewidth]{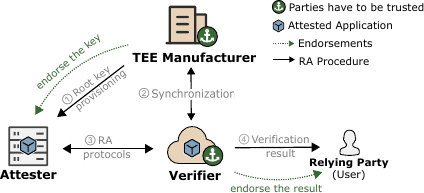}
  \caption{A typical TEE RA workflow.}
  \label{fig:RA_illustrate}
\end{figure}

\textbf{Remote Attestation (RA)} is an essential component in modern TEEs. It relies on trusted measurement and cryptographic primitives to extend the chain of trust between remote entities. Figure \ref{fig:RA_illustrate} depicts a simplified RA workflow standardized in RFC9334 \cite{RATSrfc}. 


We further explain the terminology \cite{RATSrfc} as follows.

\begin{itemize}[itemsep=0pt, parsep=0pt, topsep=2pt]
  \item \textit{TEE Manufacturer} ($\manu$) provisions RoT in the TEE devices, \eg, root key injection (\ding{192}). Users have to trust Intel \cite{costan2016intel}, Apple \cite{appletee}, Samsung \cite{samsungtee} and other semiconductor giants to securely embed a private root key.
  \item $\manu$ synchronizes necessary parameters (\ding{193}) with \textit{Verifier} ($\vrf$), such as device specifications and public keys. $\vrf$ is usually a cloud verification service, like Azure \cite{azurera}, serving as a proxy between the users and the TEEs. It initiates and conducts the RA procedure with the attester.
  \item \textit{Attester} ($\att$) is an application deployed in the TEE. For example, in SGX, $\att$ obtains its binary digest by \texttt{MRENCLAVE}, which is signed by an attestation key derived from the root key. It then produces an encrypted RA report for $\vrf$ to verify (\ding{194}). This procedure can be mutual \cite{chenMAGEMutualAttestation2022} in some distributed systems where $\vrf$ is also attested.
  \item \textit{Relying Party} ($\srpr$) is the actual user of RA who requests $\vrf$ for attestation. It can decide whether to trust $\att$ as authentic based on the verification results (\ding{195}). 
\end{itemize}


RA is the initial procedure occurring before TEE executes actual applications. Users should use RA to verify whether their deployed code is authentic before launching it.

\subsection{Physically Unclonable Functions (PUF)}

PUF, developed from the physical one-way function \cite{pappuPhysicalOneWayFunctions2002}, is a hardware primitive that randomly maps a given challenge to a response. They utilizes minor environmental variations in manufacturing and incorporate physical randomness into its structures. Therefore, each PUF instance has a unique structure with a distinct set of challenge-response pairs (CRPs). In this work, PUF is treated as a stable physical random function \cite{RuhrmairPUFSP}.

\noindent\textbf{PUF in the real-world}. Currently, famous MCU and FPGA manufacturers have all introduced their commercialized PUFs \cite{nxplpcpuf, xilinxpuf}. There are also several PUF-specialized companies including Verayo, Intrinsic-ID \cite{intrinsicidpuf}, ICTK \cite{ictkpuf}, and PUFtech. PUF has already been integrated in various real-world products, such as Intel SGX \cite{intelpuf}, IoT devices \cite{pufiotrot}, medical wearables \cite{pufmedwearable} and USIM \cite{pufusim}. With the standardization efforts \cite{PUFISOP1, PUFISOP2}, PUF is experiencing widespread deployment and becoming a basic component in hardware ecosystems.

The most common usage of PUF today is root key generation. A PUF response can be a secret seed to derive a recoverable root key \cite{costan2016sanctum,RomanPUFHash}. However, this usage does not provide stronger security guarantees because the RA procedure still becomes compromised if the key is revealed \cite{sgaxe,VoltJockeyQiuWLQ19}. A huge gap exists between the rapidly growing PUF industry and less developed PUF usages.




%





\section{Motivation}\label{sec:motivation}

\subsection{Lessons Learned from Real-world TEE RA Designs}\label{sec:lesson_learned}


Currently, nearly all real-world vendors \cite{costan2016intel,azurera, awsra,googlera} have adopted the workflow presented in \cref{sec:tee_ra_background} since it relies on a straightforward trust model. We then present three mainstream commercial designs and analyze their problems. 


$\bullet$\ Intel IAS-based RA \cite{costan2016intel}: An attesting application obtains its binary measurement by \texttt{MRENCLAVE}. It wraps the measurement, a signature signed by an Intel-provisioned key and other data to form an encrypted report. This report can only be decrypted and verified by Intel Attestation Service (IAS).

$\bullet$\ Intel DCAP-based RA \cite{scarlataSupportingThirdParty}: To avoid the IAS-centric procedure, Intel proposed Data Center Attestation Primitives (DCAP) which offloads the report verification to a third party by issuing Provisioning Certification Key (PCK) certificates. However, users still rely on Intel's root key, and only one (non-Intel) verifier can execute the verification.

$\bullet$\ Microsoft Azure RA \cite{azurera}: Similar to IAS-based RA, only the Azure Attestation Service can verify the measurements in this design. The main difference is the replacement of IAS with another vendor-specific service. $\srpr$ has to trust Azure ($\vrf$) to return the right results. Other cloud vendors (\eg, Google \cite{googlera}, AWS \cite{awsra}) adopt a similar architecture. 


In these examples, as long as $\vrf$ confirms $\att$'s report that is endorsed by $\manu$, the chain of trust is established. $\srpr$ has no alternative but to believe that $\vrf$ has faithfully executed the verification and that the TEE root key is secure. However, powerful attackers \cite{kocher2020spectre,Lipp0G0HFHMKGYH18Meltdown,schaik2021cacheout,PLATYPUSLippKOSECG21} can generate fake RA reports and verification results once they leak the key or corrupt $\vrf$. Then, others would not be able to find the compromises of TEEs if the trust in RA procedure has been completely broken. Also, these RA designs themselves cannot provide reliable functionalities. How they can be used is entirely up to the underlying cloud infrastructure. They serve merely as only add-ons to the deployed cloud services for TEEs. 




\noindent\textbf{Lessons Learned}: We conclude the following lessons, \ie, the motivation of this work, from the architectural weaknesses of these real-world RA designs.

\begin{itemize}[itemsep=0pt, parsep=0pt, topsep=2pt]
    \item Current designs rely on $\manu$ to provision a single centralized root key as the RoT. Once the key is compromised, attackers can tamper with the deployed applications and generate fake signed RA reports at will. (Requiring \textbf{R1})
    \item Only one $\vrf$ can execute the verification and $\srpr$ must accept whatever $\vrf$ returns. The results cannot be re-verified or audited to ensure that $\vrf$ has honestly checked the attested applications. (Requiring \textbf{R2})
    \item The usability of current designs is entirely up to the underlying cloud infrastructure. They cannot provide reliable attestation services without a stable (cloud) environment. (Requiring \textbf{R3})
\end{itemize}







\subsection{Related Works and Design Motivation}\label{sec:motivation_related_work}

Driven by these real-world needs, we now investigate the related works and their limitations. Further, we will clarify how \janus\ is motivated to adopt PUF and smart contract to satisfy the three security requirements. The works \cite{nunesVRASEDVerifiedHardware2019, deoliveiranunesTOCTOUProblemRemote2021,APEX,surminskiRealSWATTRemoteSoftwarebased2021,sunOATAttestingOperation2020,wangARIAttestationRealtime, kuangFeSAAutomaticFederated2022, carpentLightweightSwarmAttestation2017, kohnhauser2019practical} that focus on other problems will not be discussed.

\subsubsection{Stronger TEE RoT for RA.} 

MATEE \cite{galanouMATEEMultimodalAttestation2022} introduces a TPM for SGX to offer an independent attestation key pair as a backup RoT. Costan \etal \cite{costan2016sanctum} and Roman \etal \cite{RomanPUFHash} use PUF to generate the root key for RA. However, the system remains vulnerable if the key is deduced by carefully designed attacks \cite{sgaxe,schaik2021cacheout,VoltJockeyQiuWLQ19,LVISPAttestation,ZombieLoad,TPMFAIL,jacob2023faultpm}, no matter how it is sealed \cite{brickell2004direct, camenisch2016universally, camenischOneTPMBind2017} or derived. The trust on a digital key string does not establish a stronger RoT for TEE RA. However, we notice that the potential of PUF has not been fully explored.


\noindent\textbf{Design motivation for R1}. The security of PUF sources from its unique physical structure. Unlike the TEE manufacturers, who have full control of the key provision, even PUF manufacturers cannot predict how a PUF instance is produced. Therefore, PUF offers physics-based independent and unbiased trust, which cannot be realized by key-based RoT. Moreover, it acts as a physical random function to provide integrity and authenticity for the input, thereby realizing measuring and ``signing" simultaneously for RA. Hence, PUF itself, instead of a PUF-derived key, can be a stronger intrinsic RoT in TEE RA to effectively decentralize the trust in the root key.

PUF has emerged as a critical security primitive in numerous applications \cite{chatterjeeBuildingPufBased2019,umarProvableSecureIdentityBased2021,qureshiPUFRAKEPUFBasedRobust2022,zhengPUFbasedMutualAuthentication2022, xiaolinSPEAR}. It also has been introduced into RA by several works \cite{schulzShortPaperLightweight2011,kongPUFattEmbeddedPlatform2014,PReFeRPUFAtt,javaidDefiningTrustIoT2020,aman2020hatt}. In PUFAtt \cite{kongPUFattEmbeddedPlatform2014}, PReFER \cite{PReFeRPUFAtt} and \cite{schulzShortPaperLightweight2011}, PUF-based measurements are verified using emulated PUF models. The system can be breached once the model is leaked by attackers. Some studies \cite{PUFRASurveyCSP, AkramMMPUF} use PUF to directly provide physical bindings for attested measurements. This approach seems similar to \janus\ since PUF is simple in use. However, we argue that our objectives differ. Compared with other works, we further observe that \textit{PUF instantiates a way to establish independent and genuine trust without human interventions}. Based on this deeper insight, we fully harness PUF's potential and address the fundamental trust crisis of TEE RA.

\subsubsection{Decentralized Verification Functions}

Intel DCAP-based RA \cite{scarlataSupportingThirdParty} and OPERA \cite{chenOPERAOpenRemote2019} delegate the authority of Intel to a third party by customized certificates, allowing it to deploy the RA verification service. This approach only migrates the trust from one centralized party to another. In fact, blockchain and smart contract are common effective techniques for achieving decentralization. TM-Coin \cite{parkTMCoinTrustworthyManagement2017}, BARRET \cite{bampatsikosBARRETTBlockchAinRegulated2019}, PERMANENT \cite{ankergardPERMANENTPubliclyVerifiable2022}, SCRAPS \cite{petziSCRAPSScalableCollective2022}, CONFIDE \cite{CONFIDESigmod} and others \cite{SGXonerated,SoKTEESmartContract22} have used them to realize the on-chain decentralized verification of RA reports. However, their contracts only enable signature verification and may impose computational overhead \cite{chenOPERAOpenRemote2019} for distributed attestations. And the historical results cannot be re-verified or audited. 

\noindent\textbf{Design Motivation for R2.} Considering the limitations of these works, we are motivated to build a more transparent procedure with extensive on-chain verification functions. We continue to use smart contract since its decentralized nature is especially suitable to satisfy R2. Compared with others, \janus\ has a more open on-chain verification procedure. We treat smart contract as a ``public bulletin board", enabling not only multi-party execution but also the preservation of historical states and results. Except for on-chain signature verification, both aspects have not been well leveraged in previous works.  


\subsubsection{Resilient RA Procedure} As mentioned earlier, the extra attestation key in MATEE \cite{galanouMATEEMultimodalAttestation2022} can cooperate with the original SGX RoT to enable the multimodal attestation. Except that, it appears that current designs \cite{aberaDIATDataIntegrity2019,Ibrahim16DARPAresilient} do not focus on providing built-in resilience against system failures such as network outages, computing exhaustion, etc., which is the fundamental service requirement for RA.

\begin{figure}[htbp]
    \centering
    \includegraphics[width=0.9\linewidth]{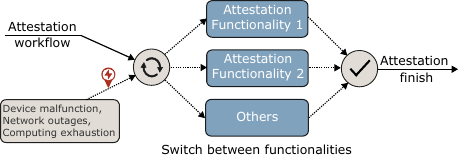}
    \caption{Illustration of attestation switch mechanism}
    \label{fig:switch_illustrate}
\end{figure}

\noindent\textbf{Design Motivation for R3.} Motivated by this significant gap, we argue that the key to realize resilience is to create redundancy in an RA design. We can develop multiple attestation functionalities. They can be activated by a switch mechanism to offer RA services collaboratively, as shown in Fig \ref{fig:switch_illustrate}. Users can choose the suitable way of performing RA in response to common system failures or customized needs, thereby ensuring the resilience. When the cloud infrastructure is limited or damaged, an RA design can mitigate the availability issues, thus continually ensuring the integrity of TEEs.




\section{\janus\ Overview}\label{sec:janus_overview}


\subsection{Threat Model and Assumption}\label{sec:threat_model}



In this work, $\att$s are assumed to be deployed in TEE platforms with PUF embedded. We use the mutual RA assumption where $\vrf$ and $\att$ are attested with each other. The TEEs that running $\vrf$s do not necessarily have PUF. We consider an adversary $\adver$ who can launch attacks against $\att, \vrf$. Its goal is to fool the attestation procedure or tamper with the results. In our threat model, $\att, \vrf$ are honest and $\adver$ can:





\begin{itemize}[itemsep=0pt, parsep=0pt, topsep=2pt]
    \item[\textbf{A1}] (\textbf{Channel attacks}) Launch man-in-the-middle (MiTM) attacks to eavesdrop, intercept, modify and replay messages between $\att$ and $\vrf$ in the open channel.
    \item[\textbf{A2}] (\textbf{DoS attacks on $\att$}) Sending massive erroneous attestation requests to an legitimate attested device to exhaust its resources, making it unavailable to handle valid requests.
    \item[\textbf{A3}] (\textbf{Physical probing}) Crack the attested devices and deduce the private keys through semi-invasive physical manners like side-channel analysis or external storage read-out.
    \item[\textbf{A4}] (\textbf{Invasive Chip delayering}) Use invasive and destructive measures like Focused Ion Beam (FIB) and Laser Voltage Probing (LVP) to mill and delayer the device chip, and then physically extract the fused keys in TEE.
    \item[\textbf{A5}] (\textbf{PUF modeling attacks}) Collect raw PUF CRPs from the communication channels during the online attestation. $\adver$ can use them to train a PUF model by machine learning to output new valid CRPs.
\end{itemize}


These attacks may help $\adver$ generate signed measurements or fool the verification procedure. Besides, $\adver$ is allowed to corrupt the verifiers to change the results. 





\begin{itemize}[itemsep=0pt, parsep=0pt, topsep=2pt]
    \item[\textbf{A6}] (\textbf{Rogue verifiers}) $\adver$ corrupts one or more legitimate $\vrf$(s) to let it/them output the verification results as $\adver$'s wishes, thereby making wrong claims on attested applications.
\end{itemize}

Compared with other works, \janus\ aims to counter the strongest possible adversary that combines all attacks in the threat model.   



\subsection{System Architecture}\label{sec:system_architecture}

\janus\ is a TEE RA system with end-to-end trusted procedure. It has a two-fold architecture with distinct RA functionalities, hence the name. For one, to address the untrusted platform limitation and satisfy \textbf{R1}, \janus\ provides PUF-based mutual attestation protocols built on the double RoTs (\ie, PUF and TEE root key). For another, to address the untrusted verifier limitation and satisfy \textbf{R2}, \janus\ provides smart contract-based verification functions that enable batch attestation and flexible result audits. The two parts together form the off-chain and on-chain RA function of \janus.


To satisfy \textbf{R3} and realize resilience for \janus, we let the two parts to collaborate by a smart contract-based switch mechanism. It yields greater openness and offers automation to switch between off-chain and on-chain attestation (\cref{sec:att_switch}). 


\begin{figure*}[htbp]
    \centering
    \includegraphics[width=\linewidth]{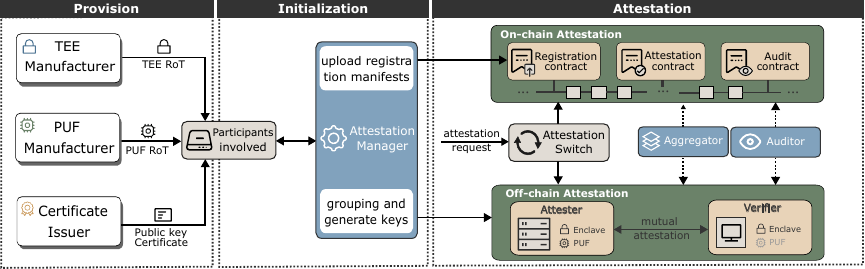}
    \caption{Architecture and workflow of \janus}
    \label{fig:janus_overview}
\end{figure*}

\noindent\textbf{Roles}: Except for $\manu, \att, \vrf$, we define the following roles for participants in \janus:


\begin{itemize}[itemsep=0pt, parsep=0pt, topsep=2pt]
    \item \textit{Attestation Manager}: A trusted third-party (TTP) service for bootstrapping the system. It does not appear in the online attestation after the initialization.
    \item \textit{Aggregator}: Resource-rich $\att$ or $\vrf$ who voluntarily aggregates blockchain transactions for others. 
    \item \textit{Auditor}: Attested legitimate $\att$ or $\vrf$ who will audit the attestation results of requested RA sessions.
\end{itemize}

The behaviors and output Attestation Manager can be verified publicly (\cref{sec:system_initialization}). It can be realized by any global service providers, thus not hurting the decentralization in practice. 

\noindent\textbf{Workflow}: We present the workflow of \janus\ in Figure \ref{fig:janus_overview}. It has three phases: Provision, Initialization, and Attestation.

\begin{itemize}[itemsep=0pt, parsep=0pt, topsep=2pt]

    \item \textit{Provision} (offline): The TEE devices are provisioned with secret materials like device-specific root keys. Also, PUF instances will be embedded into the devices as an intrinsic RoT to provide independent trust.

    \item \textit{Initialization} (offline): $\att$ and $\vrf$ and other participants first set up a consortium blockchain. Then Attestation Manager groups $\att$ and $\vrf$, and uploads registration manifests to the blockchain. Details can be found in \cref{sec:system_initialization}
    
    \item \textit{Attestation} (online): For a complete RA session in \janus, $\vrf$ and $\att$ first choose the way of attestation by switch. Then $\vrf$ sends a challenge message to start the off-chain (\cref{sec:off_chain_att}) or on-chain attestation \cref{sec:off_chain_att}. The result will be determined and may be audited in the future.
\end{itemize}

\subsection{Design Challenges}\label{sec:design_challenge}


\noindent\textbf{C1: Maintaining scalability in an end-to-end (E2E) protocol.}\label{challenge1} Common approaches of establishing secure E2E links between two parties use heavyweight primitives (TLS) or redundant interactions (two-party computation), limited in efficiency and scalability. This challenge is particularly tricky in our work since PUF is a hardware-unique primitive that each instance has a unique CRP domain. Pre-sharing each ones' CRPs would bring tremendous overhead instead.




\noindent\textbf{C2: Avoiding attestation data leakage on the blockchain.}\label{challenge2} The blockchain used here may allow rogue verifiers (\textbf{A6}) to probe data on-chain and recover the attestation messages in other RA sessions or deduce others' secrets. For example, raw measurements or PUF CRPs must not be stored on-chain since rogue verifiers may directly use them to perform MITM attacks (\textbf{A1}).




\noindent\textbf{C3: Snapshotting an RA session.}\label{challenge3} To enable the audits of attestation results, uploading all states and output of RA sessions to the blockchain is a straightforward approach. However, it is infeasible since this causes excessive storage overhead and latency on-chain. Therefore, to keep the scale of blockchain compact and functionally focused, we need a lightweight and secure structure to properly summarize RA sessions, \ie, snapshots. Verifying the snapshots would be equivalent to verifying each step of the protocol. 



\noindent\textbf{C4: Realizing automation for attestation switch.}\label{challenge4} The two RA functionalities provided by \janus\ should be closely merged with each other to achieve resilience for various applications and situations. The switch mechanism must be self-instructing and interconnect with each part, without requiring global operators to manage the RA procedures. 



\section{\janus\ Designs}\label{sec:janus_design}

We present the detailed design of \janus\ in this section. As described in \cref{sec:system_architecture}, we realize two attestation functionalities, \ie, off-chain and on-chain, and an automated switch mechanism to combine the two. The notations used through this paper are given as follows.



\noindent\textbf{Notations}: Let $\bin^{*}$ be the set of all bit strings and $\emptyset$ be an empty set. For any $n \in \mathbb{N}$, $\bin^{n}$ is the set of all $n$-bit strings. $X \sample \bin^{n}$ means the random selection of a string $X$ from $\bin^{n}$. For some $X \in \bin^{n}$, $|X|$ denotes the length of $X$. For some $X, Y \in \bin^{n}$, $X || Y$ denotes their concatenation, and $X \oplus Y$ denotes their bitwise XOR result. The lengths of PUF challenges and responses are set to be the same in this study, denoted by $n$ unless otherwise specified. $H_{K}(\cdot)$ denotes a keyed hash function.


\subsection{System Initialization}\label{sec:system_initialization}

The offline initialization is conducted by the Attestation Manager, who helps generate IDs and protocol keys. The Manager is \textit{not} involved in the online attestation. 



\ding{172} \textbf{Setup and Grouping.} The participants first create their blockchain accounts, \ie a key pair $(\pk, \sk)$ with personal address $\mathsf{addr}=H(\pk)$. Then the Manager does the following:
\begin{itemize}[itemsep=0pt, parsep=0pt, topsep=2pt]
    \item \underline{Grouping}: Manager organizes the participants in \janus\ by their locations or functions. For example, the attested applications on the same device can be naturally assigned in the same group (group id $\gid$). The use of grouping will be further shown in \ding{173}.
    \item \underline{Attesters}: Let $\aid=\mathsf{addr}||\mathsf{DSN}||\gid$. $\mathsf{DSN}$ (Device Serial Number) identifies on which device $\att_{\aid}$ is deployed.
    \item \underline{Verifiers}: Let $\vid=\mathsf{addr}||\mathsf{VON}||\gid$ where $\mathsf{SON}$ (Verifier Owner Number) specifies which owner $\vrf_{\vid}$ belongs to. $\vid$ is similar to $\aid$, which eases the identity switch in the mutual attestation.
\end{itemize}




\ding{173} \textbf{Protocol keys generation.} Then Manager helps generate the protocol key, including a group key (unique for each group) and a communication key (unique for each one).
\begin{itemize}[itemsep=0pt, parsep=0pt, topsep=2pt]
    \item \underline{Attesters:} Each $\att_{\aid}$ generates an initial PUF challenge $C_{\mathsf{init}}$, which can be chosen arbitrarily and made public. Then $\att_{\aid}$ uses PUF to get $R_{\aid}=\puf{}\ (C_{\mathsf{init}})$ and $\mli{MR}_{\aid}=\puf{}\ (R_{\aid})$, and submits $R_{\aid}$ to the Manager. The Manager calculates and returns the group key $K_{\gid}=\oplus_{i=1}^{N}R_{i}$ for a group $G_{\gid}$ with $N$ attesters. $\att_{\aid}$ now calculates its communication key $\mli{MK}_{\aid}=K_{\gid}\oplus R_{\aid} \oplus \mli{MR}_{\aid}$.
    \item  \underline{Verifiers:} Each $\vrf_{\vid}$ randomly selects a communication key $s_{\vid}\sample\bin^{n}$, then submits $s_{\vid}$ to the Manager and receives their group key $S_{\gid}=\oplus_{j=1}^{M}s_{j}$. 
    \item  \underline{Manager:} For each communication key, it encrypts the key using all group keys to compute the handshake materials $\mathcal{H}$. For example, for $\mli{MK}_{i}$ of $\att_{i}$, $\mathcal{H}_{i}=\{\dots,\mathsf{Enc}_{K_{j}}(\mli{MK}_{i}),\mathsf{Enc}_{K_{i}}(\mli{MK}_{i}),\dots\mathsf{Enc}_{S_{k}}(\mli{MK}_{i}),\dots\}$. All $\mathcal{H}$ will be uploaded to the blockchain in \ding{174}.
    This pre-computation can address \textbf{\hyperref[challenge1]{C1}} and thus significantly improve the E2E scalability of \janus.
    
\end{itemize}





Finally, $\att_{\aid}$ stores $(C_{\mathsf{init}}, \mli{MK}_{\aid})$ and $\vrf_{\vid}$ stores $(s_{\vid}, S_{\gid})$. $\att_{\aid}$ does not need to store the group key since they can recover $K_{\gid}=\mli{MK}_{\aid}\oplus \mli{MR}_{\aid} \oplus R_{\aid}$ by PUF. Note that the communication keys are designed to be known by the other participant in the protocol (\cref{sec:remote_attestation}). Our PUF-based designs ensures that an adversary obtaining several communication keys cannot derive any group key, eliminating the usage of secure storage (seen in \cref{sec:protocol_summary}).


\ding{174} \textbf{Registration to the blockchain.} The Manager uploads registration manifests to the blockchain, including $\mathcal{H}$, device configurations, and measurements. The configurations would contain CPU SVN versions, TCB sizes, and others. The measurements fall into two types here:
\begin{itemize}[itemsep=0pt, parsep=0pt, topsep=2pt]
    \item \underline{Traditional Measurement}: For verifiers in mutual RA, their measurements are traditional enclave measurements $M$, \eg, hash digests of the application's meta-data \cite{costan2016intel}.
    \item \underline{PUF-based Nested Measurement}: The measurements $RM$ of attesters are obtained using both PUF and TEE. Specifically, we let $RM=\puf{}\ (M)$ where $M$ is the original enclave measurement. 
\end{itemize}

%


The Manager would upload $(\pid, H(RM_{\aid}||\aid||\pid))$ or $H(M_{\vid}||\vid)$ instead of raw measurements to prevent the sloppy attesters in \textbf{challenge \hyperref[challenge2]{C2}}. The registration contract handling these manifests is given in \cref{sec:on_chain_attestation}.


\subsection{PUF-based Attestation Protocols}\label{sec:off_chain_att}


To satisfy \textbf{R1}, we use PUF as an intrinsic RoT in TEE to generate physically trusted measurements. The designed attestation protocols form the off-chain RA function of \janus.






\subsubsection{Local attestation}\label{sec:local_attestation}

\janus\ supports the local attestation (LA) like Intel SGX RA \cite{costan2016intel}. Multiple $\att$s can exist as seperate functions on the same TEE device. LA allows them to attest each other before interacting with a remote $\vrf$.

\begin{figure}[!htbp]
    \centering
    \includegraphics[width=\linewidth]{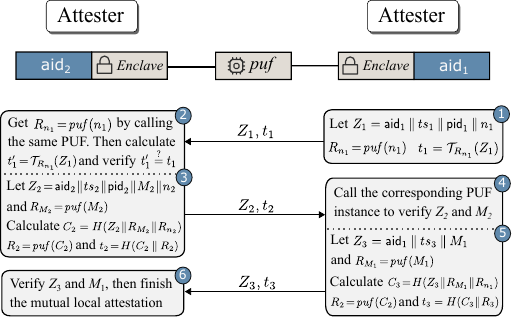}
    \caption{The off-chain local attestation protocol of \janus}
    \captionsetup{font=small}
    \caption*{
    \ding{172} $\att_{\aid_{1}}$ uses $\aid_{1}$, timestamp $ts_{1}$, local PUF id $\pid_{1}$ and a nonce $n_{1}\in\bin^{n}$ to form $Z_{1}$. Then it invokes PUF to get $R_{n_{1}}$. A MAC algorithm $\mathcal{T}$ like HMAC \cite{hmacbellare} is used with $R_{n_{1}}$ being the key. \\
    \ding{173} $\att_{\aid_{2}}$ reproduces $R_{n_{1}}$ by using $n_{1}$ as the PUF challenge. $(Z_{1}, t_{1})$ is considered valid only if $\att_{\aid_{2}}$ gets the same $t_{1}$. \\
    \ding{174} After verifying $(Z_{1}, t_{1})$, $\att_{\aid_{2}}$ uses a different PUF instance $\pid_{2}$ to get $R_{M_{2}}$. It obtains $C_{2}, R_{2}=\puf{}(C_{2})$ and gets $t_{2}$. \\
    \ding{175} $\att_{\aid_{1}}$ uses the specified PUF instance to get $R_{n_{2}}$ and re-calculated $C_{2}, R_{2}$, thereby verifying $(Z_{2}, t_{2})$. \\
    \ding{176} $\att_{\aid_{1}}$ follows the same steps in \ding{174} to generate $(Z_{3}, t_{3})$. Its measurement $M_{1}$ is also contained in $Z_{3}$ to attest $\att_{\aid_{2}}$. \\
    \ding{177} If the verification passes for both enclaves, it proves that they indeed exist on the same TEE platform.
    }
    \label{fig:local_attestation}
\end{figure}

In Figure \ref{fig:local_attestation}, two attesters $\att_{\aid_{1}}, \att_{\aid_{2}}$ control different enclaves and can both invoke the device's PUF instances. We let $\att_{\aid_{1}}$ initiate the session and generates the challenge $(Z_{1}, t_{1})$ by step \ding{172}. When receiving $(Z_{1}, t_{1})$ from an untrusted local channel, $\att_{\aid_{2}}$ invokes the same PUF to verify the tag. The operations in step \ding{174} and \ding{176} provides PUF-based integrity and authenticity to the tag generation. Our protocol avoids leaking PUF responses to the channel, or adversaries can collect raw CRPs to perform PUF modeling attacks (\textbf{A5}).


\subsubsection{Remote attestation}\label{sec:remote_attestation}

\begin{figure}[htbp]
    \centering
    \includegraphics[width=\linewidth]{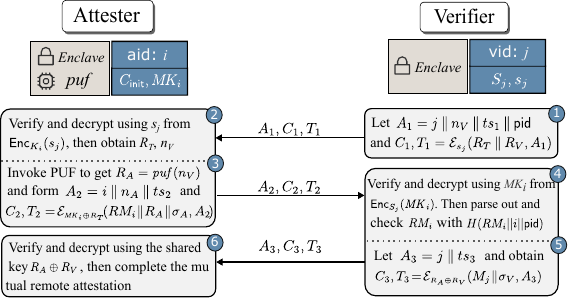}
    \caption{The off-chain remote attestation protocol of \janus}
    \label{fig:remote_attestation}
\end{figure}

This protocol is the core of off-chain RA function of \janus. As shown in Figure \ref{fig:remote_attestation}, an $\att$ and a $\vrf$ is mutually attested through this three-round protocol.




Before starting the RA session, the verifier $\vrf_{j}$ retrieves $\mathsf{Enc}_{S_{j}}(\mli{MK}_{i})$ and $(\pid, H(RM_{i}||i||\pid))$ from the blockchain. Then it initiates the RA session as follows.

\noindent\textbf{Sending remote attestation challenge.} $\vrf_{j}$ first constructs the challenge message $(A_{1}, C_{1}, T_{1})$:

\begin{itemize}[itemsep=0pt, parsep=0pt, topsep=2pt]
    \item [\ding{172}] \uline{$(A_{1}, C_{1}, T_{1})$}:  $\vrf_{j}$ forms $A_{1}$ and samples $R_{T}, R_{V}\sample\bin^{n}$ where $R_{T}$ is a temporary random string and $R_{V}$ is a part of the shared key. Now $\vrf_{j}$ uses an authenticated encryption (AE) algorithm $\mathcal{E}$ to encrypt $R_{T} || R_{V}$ with $A_{1}$ as the associated data (AD) and $s_{j}$ as the key. Thus, $\vrf_{j}$ gets the ciphertext $C_{1}$ and the tag $T_{1}$, then sends $(A_{1}, C_{1}, T_{1})$ to the attester.
\end{itemize}


\noindent\textbf{Responding with signed measurement.} Upon receiving $(A_{1}, C_{1}, T_{1})$, $\att_{i}$ learns the $\vid_j$ and retrieves $\mathsf{Enc}_{K_{i}}(s_{j})$, $H(M_{j}||j)$ from the blockchain. 

\begin{itemize}[itemsep=0pt, parsep=0pt, topsep=2pt]
    \item [\ding{173}] \uline{verify $(A_{1}, C_{1}, T_{1})$}:  $\att_{i}$ gets $R_{i}, MR_{i}$ by PUF and recovers $K_{i}=\mli{MK_{i}}\oplus \mli{MR}_{i} \oplus R_{i}$. It uses the decryption algorithm $\mathcal{D}$ to verify $(A_{1}, C_{1})$ and obtain $R_{T}, R_{V}$.
    \item [\ding{174}] \uline{$(A_{2}, C_{2}, T_{2})$}: After $\att_{i}$ verifies the challenge, it generates the nested measurement $RM_{i}=\puf\ (M_{i})$. Then it uses original TEE attestation key to sign $RM_{i} || R_{A}$ where $R_{A}=\puf_{\pid}\ (n_{V})$ is another part of the shared session key. The signature $\sigma_{A}$ is then appended after $RM_{i} || R_{A}$ and they are together encrypted by $\mathcal{E}$. Finally, $\att_{i}$ sends $(A_{2}, C_{2}, T_{2})$ back to $\vrf_{j}$.
\end{itemize}


\noindent\textbf{Verifying with response.} $\vrf_{j}$ decrypts $\mathsf{Enc}_{S_{j}}(\mli{MK}_{i})$ to get $\mli{MK}_{i}$ so that it can calculate $\mli{MK}_{i} \oplus R_{T}$. 

\begin{itemize}[itemsep=0pt, parsep=0pt, topsep=2pt]
    \item [\ding{175}] \uline{verify the measurement}: $\vrf_{j}$ uses the certificate $\mathsf{Cert}_{i}$ to verify $\sigma_{A}$. If the verification passes, $\vrf_{j}$ parses out $RM_{i}$ to re-calculate $H(RM_{i}||i||\pid)$ and compares it with the one on-chain. If the measurement is valid, then $\vrf_{j}$ completes the attestation with $\att_{i}$.
    \item [\ding{176}] \uline{$(A_{3}, C_{3}, T_{3})$}: $\vrf_{j}$ now can obtain the shared session key $SK=R_{A}\oplus R_{V}$. Then it forms $A_{3}$ and generates its enclave measurement $M_{j}$. It calls $\mathcal{E}$ to encrypt $M_{j}$ along $\sigma_{V}$ using the shared key $SK$.
\end{itemize}

In step \ding{177}, $\att_{i}$ can verify $(A_{3}, C_{3}, T_{3})$ and the measurement $M_{j}$ using the same procedure in \ding{175}. Then it finishes the attestation with $\vrf_{j}$ and obtains $SK=R_{A}\oplus R_{T}$.

\subsubsection{Protocol Summary}\label{sec:protocol_summary}

We summarize the above attestation protocols and give the following remarks.

\begin{itemize}
    \item \textbf{Intrinsic RoTs}: With PUF being an extra intrinsic RoT, $\adver$ would fail the attestation even if it knows the TEE root keys. This decentralized trust can offer stronger security guarantees for TEE RA and succinctly addresses the untrusted TEE platform limitaion. 
    \item \textbf{Performance Tradeoff}: By the pre-computation of handshake materials, our protocol achieves lightweight and scalable communication at the cost of single query to blockchain. Participants can periodically cache the on-chain entries, minimizing the latency of the online query.
    \item \textbf{Key Protection}: The designed key structures of $\att$ use PUF response $\mli{MR}_{\aid}$ to mask the secret $R_{\aid}$. Then even if $\mli{MK}_{\aid}$ is leaked, $\adver$ will learn nothing about $R_{\aid}$ and $K_{\gid}$ (\textbf{A3}).
\end{itemize}
\noindent





\subsection{Smart Contracts for Attestation}\label{sec:on_chain_attestation}

We design three smart contracts in \janus\ to realize participant registration, on-chain attestation, and attestation audit. We first introduce the registration contract, which is required by both off-chain and on-chain attestation.

\subsubsection{Registration Contract}\label{sec:registration_contract}

As mentioned in \cref{sec:system_initialization}, thi contract is used to store registration manifests. It manipulates the blockchain as an immutable distributed database $\mathcal{L}$ through key-value pairs and offers two functionalities:

\noindent\ding{172}\ \underline{\textsc{Upload manifests}}. The Attestation Manager can upload devices' measurements, encrypted key materials and other data through this contract. The contract stores them with unique blockchain addresses as indexes in $\mathcal{L}$.

\noindent\ding{173}\ \underline{\textsc{Retrieve on-chain data}}. The blockchain nodes (initialized devices) are allowed to retrieve any data stored on-chain. They can use a specified address to query the block entry in $\mathcal{L}$ without direct interactions with the contract. 

The registration contract avoids the risk of a single point of failure caused by a central server, thereby offering communication scalability for RA in large-scale network.


\subsubsection{Attestation Contract}\label{sec:attestation_contract}

This contract realizes on-chain attestation for \janus. It involves multiple verifiers in a single verification execution. As a result, both the execution and results of RA verification would be publicly aware and admitted. 

The workflow of this contract is described as follows. The verifier generates a request transaction to trigger the attester, \eg, changing the state nonce $n$ under a given request blockchain address which $\att_{\aid}$ periodically queries. When $\att_{\aid}$ learns of a coming request (the nonce being changed), it generates $RM_{\aid}, \sigma_{A}$, and calculates $C=\mathsf{Enc}_{\mli{MK}_{\aid}}(n||RM_{\aid}||\sigma_{A})$. It wraps $C$ in a valid transaction and submits it to a nearby Aggregator (\cref{sec:system_architecture}). Aggregators will wrap and submit received transactions in a batch to the attestation contract. Main functions of this contract are formalized in Algorithm \ref{alg:attestation_contract}.



\begin{algorithm} 
    \caption{Smart contract for on-chain attestation}\label{alg:attestation_contract}
    \SetNoFillComment
    \Input{transaction list $l$}
    \Output{trust state $st$}
    \nosemic\nonl\qquad\qquad\qquad\enskip\underline{\textsc{Batch attestation}}\;
    \For{$\mathrm{a\ transaction}\ tx \in l$}{
        \If{$tx\mathsf{.signature}$ is valid}{
            $st\leftarrow$\textsc{On-chain attestation}($tx$)\;
            \ShowLn\Output{$st$}
        }
    }
    \nosemic\nonl\qquad\qquad\qquad\underline{\textsc{On-chain attestation}}\;
    $c \leftarrow\mathsf{Enc}_{S_{g}}(MK_{\aid}),\enskip t\leftarrow H(RM_{\aid}||\aid||\pid),\enskip \pid$\; \label{alg:attcont_7}
    \tcc{retrieve them from blockchain}
    $MK_{\aid}\!\leftarrow\!\mathsf{Dec}_{S_{g}}(c), C \leftarrow tx.\mathsf{payload}$ \tcp{parse tx}
    $P\leftarrow\mathsf{Dec}_{MK_{\aid}}(C)$,\quad $(RM_{\aid}^{*}, \sigma_{A}) \leftarrow \mathsf{Parse}(P)$\;
    $t^{*}\leftarrow H(RM_{\aid}^{*}||\aid||\pid)$ \;
    \If{$\sigma_{A}$ is invalid or $t \neq t^{*}$}{
        \ShowLn\Output{\textsf{untrusted}}
    }
    \ShowLn\Output{\textsf{trusted}}
\end{algorithm}

\noindent\ding{172}\ \underline{\textsc{On-chain attestation}}: As shown in line \ref{alg:attcont_7}, verifiers use $\mli{MK}_{\aid}$ to obtain the measurement $RM_{\aid}$ and the signature $\sigma_{A}$. They can use the same verification steps in \cref{sec:remote_attestation} and determine the attestation result on $\att_{\aid}$ together.

\noindent\ding{173}\ \underline{\textsc{Batch attestation}}: Verifiers can verify multiple measurements in a transaction batch at once. This improves computing resource utilization of the blockchain nodes and facilitates the on-chain verification procedure. 






\subsubsection{Audit Contract}\label{sec:audit_contract}

\begin{figure}[htbp]
    \centering
    \includegraphics[width=0.9\linewidth]{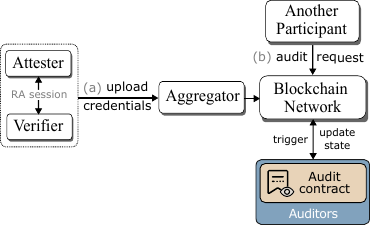}
    \caption{Workflow of on-chain audit smart contract}
    \label{fig:on_chain_audit_contract}
\end{figure}

We design an audit contract as shown in Figure \ref{fig:on_chain_audit_contract}. This contract allows other participants in \janus\ to re-verify the result of a historical RA session. It helps evaluate the device status and identify compromised verifiers or attesters in time.




As mentioned before, uploading all protocol messages for all $\att$ and $\vrf$ is not feasible (\textbf{\hyperref[challenge3]{C3}}), so we propose a new data structure, \textit{credential}, to snapshot an end-to-end RA session. Here, $\att_{\aid}$ calculates $cr_{1}=H_{K_{\gid}}(H_{s_{\vid}}(m_{1}||m_{2}||\aid||\vid))$ and $\vrf_{\vid}$ calculates $cr_{2}=H_{S_{\gid}}(H_{\mli{MK}_{\aid}}(m_{1}||m_{2}||\aid||\vid))$. A pair of credentials $(cr_{1}, cr_{2})$ ensures that both parties have executed the protocol and acquired the expected output, \ie, obtained each other's key ($K_{\gid}$ or $s_{\vid}$) and measurements ($m_{1}$ or $m_{2}$). $(cr_{1}, cr_{2})$ will be submitted to the blockchain through an Aggregator for audits. To this end, We define a new trusted entity, Auditor. Legitimate $\att$ or $\vrf$ that has been attested can become an Auditor to execute this contract in their TEEs.

Our audit contract provides three functions, as formalized in Algorithm \ref{alg:audit_contract}.





\begin{algorithm}
    \SetKwInOut{Input}{Input}\SetKwInOut{Output}{Output}
    \caption{Smart contract for attestation audit}\label{alg:audit_contract}
    \SetNoFillComment
    \Input{transaction list $l$}
    \Output{device trust rate $r$, audit state $st$}
    \nosemic\nonl\qquad\qquad\qquad\quad\underline{\textsc{Trust rating}}\;
    \If{$tx\mathsf{.signature}$ is valid}{
        $\mathsf{conf}\leftarrow\mathsf{Parse}(tx.\mathsf{payload})$\tcp{device config}
        $r\leftarrow \mathsf{Evaluate}(\mathsf{conf})$\;\label{alg:audcont_18}
        \ShowLn\Output{$r$}
    }
    \nosemic\nonl\qquad\qquad\qquad\quad\underline{\textsc{Spot check:}}\;
    $n\sample [0, l\mathsf{.size}-1], \enskip tx \leftarrow l[n]$\;
    \tcc{randomly pick one}
    \If{$tx\mathsf{.signature}$ is valid}{
        $st\leftarrow$\textsc{Attestation review}($tx$)\;
        \ShowLn\Output{$st$}
    }
    \nosemic\nonl\qquad\qquad\qquad\underline{\textsc{Attestation review}}\;
    $K_{\gid},S_{\gid}, MK_{\aid},s_{\vid},m_{1}\leftarrow RM_{\aid},m_{2}\leftarrow M_{\vid}$\;
    \tcc{retrieve them from the Manager}
    $(cr_{1}, cr_{2}) \leftarrow \mathsf{Parse}(tx.\mathsf{payload})$ \tcp{credentials}
    $cr_{1}^{*}\leftarrow H_{K_{\gid}}(H_{s_{\vid}}(m_{1}||m_{2}||\aid||\vid))$\;\label{alg:audcont_9}
    $cr_{2}^{*}\leftarrow H_{S_{\gid}}(H_{MK_{\aid}}(m_{1}||m_{2}||\aid||\vid))$\;\label{alg:audcont_10}
    \If{$cr_{1} = cr_{1}^{*}$ and $cr_{2} = cr_{2}^{*}$}{
        \ShowLn\Output{\textsf{trusted}}
    }
    \ShowLn\Output{\textsf{untrusted}}
    
\end{algorithm}

\noindent\ding{172}\ \underline{\textsc{Trust rating}}: Participants can upload their configurations to the blockchain (\cref{sec:registration_contract}). Auditors can use this contract to calculate the trust rate based on the configurations of deployed device. Trust rates can serve as indicators of whether the status of $\att$ should be auditted in time.



\noindent\ding{173}\ \underline{\textsc{Attestation review}}: Auditors can obtain keys and measurements from the Manager to verify the credentials. As shown in line \ref{alg:audcont_9}, \ref{alg:audcont_10}, they can re-calculate $(cr_{1}^{*}, cr_{2}^{*})$ and compare them with $(cr_{1}, cr_{2})$ in a submitted transaction. 


\noindent\ding{174}\ \underline{\textsc{Spot check}}: Auditing every credential can bring tremendous computation overhead. Therefore, we allow Auditors to \textit{spot check} some credentials, \ie, treat the audit results of randomly picked credentials as the results of the whole transaction list. Combined with trust rating, spot check would significantly improve audit efficiency and minimizes the risk of missing compromised participants.



The attestation and audit contract well satisfy \textbf{R2} by enabling various decentralized verification functions and providing consensual and auditable results. Every RA session in \janus\ are now publicly transparent and can be re-verified efficiently, preventing malicous verifiers from exclusively claiming wrong attestation results.




\subsection{Attestation Switch in \janus}\label{sec:att_switch}

\begin{figure}[htbp]
    \centering
    \includegraphics[width=0.85\linewidth]{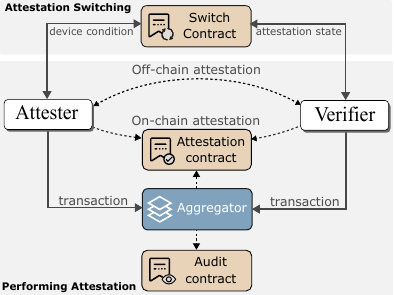}
    \caption{Holistic view of attestation switch in \janus}
    \label{fig:attestation_switch}
\end{figure}

The switch machanism aims to make two functionalities in \janus\ to collaborate with each other. As shown in Figure \ref{fig:attestation_switch}, it is instantiated by a switch smart contract and switch rules. When encountering network outages and other incidents, $\vrf$ and $\att$ can asynchronously determine how they will perform RA. No other entities or global operators are required, so $\vrf$ and $\att$ can automatically control their attestation flow, which addresses the rigid procedure limitation and satisfy (\textbf{R3}). 




\noindent\textbf{Switch contract}: This contract maintains two fields for every $\att$ to change the attestation flow: device condition (\texttt{dc}) and attestation state (\texttt{as}). Both fields take one of the values in \{\texttt{off-chain}, \texttt{on-chain}, \texttt{both}\}. Note that \texttt{dc} is only set by $\att$ and \texttt{as} is set by anyone who wants to conduct RA with $\att$, \ie, some verifier. 


\texttt{dc} indicates the way in which $\att$ prefers to conduct the attestation, \ie, off-chain protocol or on-chain smart contract. In practice, the device where $\att$ resides may experience network outages, low battery power, stressed computing resources, etc. \texttt{dc} gives $\att$ the leeway to adapt these conditions. \texttt{as}, on the other hand, indicates how $\vrf$ will initiate an RA session with $\att$. $\vrf$ first checks $\att$'s \texttt{dc}, and if \texttt{dc}=\texttt{both}, then $\vrf$ can set \texttt{as} to \texttt{on-chain} or \texttt{off-chain}. Otherwise, \texttt{as} should only be set the same as \texttt{dc}. 

For example, if $\att$ cannot establish point-to-point channels currently, it can trigger the switch contract to update \texttt{dc} to \texttt{on-chain} since the blockchain is always online. Then $\vrf$ will choose to conduct on-chain RA functions with $\att$ after it knows $\att$'s \texttt{dc}. They can change their attestation flow independently without interference of any other operators.

\noindent\textbf{Switch rules}: We now give some example switch rules to instruct the switch procedure.



\underline{\textsc{Switch to on-chain}}: If \texttt{dc} $\neq$ \texttt{off-chain}, $\vrf$ can choose on-chain smart contract-based attestation with $\att$ under the following cases:

\begin{itemize}[itemsep=0pt, parsep=0pt, topsep=0pt]
    \item $\vrf$ cannot establish direct communication with $\att$.
    \item The attestation results need to be acknowledged by at least a quorum of verifiers.
    \item $\att$'s application is deployed in a distributed manner.
\end{itemize}

$\vrf$ needs to update \texttt{as} to \texttt{on-chain} through the switch contract. Then it generates a request transaction to perform the normal on-chain attestation (\cref{sec:attestation_contract}).


\underline{\textsc{Switch to off-chain}}: If \texttt{dc}$\neq$\texttt{on-chain}, $\vrf$ can choose off-chain PUF-based attestation protocols under the following cases:
\begin{itemize}[itemsep=0pt, parsep=0pt, topsep=0pt]
    \item $\vrf$ and $\att$ only need to conduct a routine attestation (periodically executed for regular check).
    \item Higher verification efficiency is required for real-time systems like IoV and IoT.
    \item $\vrf$ needs off-chain attestation for other reasons, \eg, privacy concerns, personal needs.
\end{itemize}

These switch rules enable $\vrf$ and $\att$ to switch between on-chain or off-chain attestation seamlessly. Administrators can configure customized rules for their RA system. In the first two cases, $\vrf$ are not required to update \texttt{as}. It directly sends the challenge of step \ding{172} in \cref{sec:remote_attestation} to start the attestation. For the third case, $\vrf$ should first update \texttt{as} to \texttt{off-chain} before the attestation. Note that $\vrf$ and $\att$ are required to submit their credentials for audits afterward in all three cases.





\subsection{Use Cases}\label{sec:use_case}

\janus\ provides a more open TEE RA framework with end-to-end trusted and resilient procedure for many real-world applications. In addition to the systems such as cloud computing and IoT, which frequently utilize RA, \janus\ can be integrated into several emerging scenarios.

$\bullet$\ \textit{Function as a Service (FaaS)}. Our on-chain functions benefit distributed services like FaaS \cite{googlefaas}. FaaS allows developers to deploy their applications in small functions. Given the highly dispersed nature of code deployment, traditional RA schemes may suffer from efficiency issues as they have to attest each function individually. In \janus, distributed functions can be attested in a batch through our contract.

$\bullet$\ \textit{BeyondCorp}. Google BeyondCorp \cite{googlebeyondcorp} is a security framework that implements a zero trust model, which shifts access controls from the network perimeter to individual devices. It requires continuous monitoring to verify the status of devices before granting access. The lightweight off-chain protocols and snapshotting mechanism in \janus\ attestation can well support these needs.

$\bullet$ \textit{Trusted Internet Connections (TIC)} CISA TIC initiative \cite{CISAInternetConnections} was established to enhance the federal networks by standardizing and consolidating network connections. It requires a trust-level evaluation and auditing capability, with the aim of reducing the number of access points while maintaining network resilience. \janus\ can also offer suitable attestation functions like trust rating and resilient RA procedure, to meet TIC standards in real-world systems in real-world systems.

\section{\janus\ security}\label{sec:janus_security}

\subsection{Security Proof of PUF-based Protocols}\label{sec:security_proof}

We first formalize the PUF definition and give the idealized PUF properties as follows.

\begin{definition}(Ideal PUF)
    Let $P$ be a PUF family where $R=\puf{}(C)$ for $ \forall \puf{} \in P$ and $C, R \in \{0, 1\}^{*}$. $P$ is an ideal PUF family if $|C|=|R|=n$ and for any probabilistic polynomial-time (PPT) distinguisher $\Aa$, we have,
    \begin{itemize}[itemsep=0pt, parsep=0pt, topsep=2pt]
		\item (Pseudorandomness) $\forall \puf{} \in P, \forall C \in \{0,1\}^{n}$, 
		\begin{equation}
			\label{eq:ideal-PUF-pseudorandomness}
			\left|\mathrm{Pr}\left[\Aa^{\puf{}(C)}\Rightarrow 1\right] - \mathrm{Pr}\left[\Aa^{\$}\Rightarrow 1 \right]\right| \le \frac{1}{2^{n}}\cdot \epsilon,
		\end{equation}
		where $\epsilon$ is a negligible value related to $n$ and $\$$ denotes a random bits oracle.
		\item (Uniqueness) $\forall \puf{}_{1}, \puf_{2} \in P, \forall C \in \{0,1\}^{n}$, let $p(C)=\puf{}_{0}(C)\oplus \puf_{1}(C)$, 
		\begin{equation}
			\label{eq:ideal-PUF-uniqueness}
				\left|\mathrm{Pr}\left[\Aa^{p(C)}\Rightarrow 1\right]-\mathrm{Pr}\left[\Aa^{\$}\Rightarrow 1\right]\right| \le \frac{1}{2^{n}}\cdot \epsilon^{'}.
		\end{equation}
        \item (Unclonability) Each PUF instance has a unique physical structure. Any attempt to evaluate a pre-installed PUF instance in simulation environments such as hardware debugging would change its response generation process and turn it into a different instance.
    \end{itemize}
\end{definition}

In this work, we utilize Universally Composable (UC) framework \cite{canetti2001universally} to depict the security of our off-chain protocols in \cref{sec:local_attestation} and \cref{sec:remote_attestation}. UC aims to rigorously demonstrate the (in)distinguishability of the protocol execution in the real world and ideal world. In the ideal world, an ideal functionality $\mathcal{F}$ describes the abstracted interfaces that a protocol wishes to achieve. Therefore, we propose the first mutual attestation ideal functionality $\mathcal{F}_{att}$ in Appendix \cref{sec:security_proof_full}.

Then a Simulator $\mathcal{S}$ in the ideal world emulates the protocol execution and behaviors of the real-world adversary $\Aa$. The protocol realizes the ideal functionality if an environment machine $\mathcal{Z}$ cannot distinguish the observed output is from the real world or the ideal world. For \janus, we have the following theorem to prove,

\begin{theorem}\label{theorem1}
    If the AEAD algorithm $(\mathcal{E}, \mathcal{D})$ provides ciphertext indistinguishability under chosen ciphertext attack (IND-CCA), the hash function $H$ is collision resistant and $\puf{}$ is an ideal PUF, then the PUF-based attestation protocols of \janus\ with respect to a global database $\Ll$ UC-realizes $\mathcal{F}_{att}$ against a static adversary $\Aa$.
\end{theorem}

\begin{proof}
    The complete proof is given in the Appendix \cref{sec:security_proof_full}.
\end{proof}

\subsection{Security Analysis}\label{sec:informal_security_analysis}

We now give the security analysis against different attacks and describe how \janus\ is capable of defending against them. As shown in Table \ref{tab:threat_model}, \janus\ can resist all attacks included in the threat model (\cref{sec:threat_model}), while other designs can only partially resist them.


\begin{table}[htpb]
	\centering
    \begin{threeparttable}
    \captionsetup{width=\columnwidth}
	\caption{Security comparison of (TEE) RA schemes\label{tab:threat_model}}
	\begin{tabular}{p{2cm}<{\centering}cccccc}
		\toprule
		\textbf{Designs} & \textbf{A1}  & \textbf{A2} & \textbf{A3} & \textbf{A4} & \textbf{A6} & \textbf{A7} \\
		\midrule
        \textbf{SGX RA}\cite{costan2016intel} & \compfull & \compnone & \comppart & \compnone & \compnone & \compnone \\
        \textbf{OPERA}\cite{chenOPERAOpenRemote2019} & \compfull & \comppart & \comppart & \compnone & \compnone & \compnone \\
        \textbf{MATEE}\cite{chenMAGEMutualAttestation2022} & \compfull & \compnone & \comppart & \compnone & \compnone & \compnone \\
        \textbf{LIRA-V}\cite{shepherdLIRAVLightweightRemote2022} & \compfull & \compnone & \comppart & \compnone & \compnone & \compnone \\\textbf{SCRAPS}\cite{petziSCRAPSScalableCollective2022} & \compfull & \compfull & \comppart & \compnone & \compfull & \compfull \\
		\textbf{\janus} & \compfull & \compfull & \compfull & \compfull & \compfull & \compfull\\
		\bottomrule
	\end{tabular}
    \begin{tablenotes}
	\small 
    \item \compfull: Resisted; \comppart: Partially resisted; \compnone: not resisted;
    \end{tablenotes}
 \end{threeparttable}
\end{table}

$\bullet$\ \textit{Channel attacks}: \janus\ has adopted complete protocol design and cryptographic techniques to prevent $\adver$ from replaying, modifying, or forging any messages in an open channel. For the off-chain PUF-based protocols, the AEAD algorithm $(\mathcal{E}, \mathcal{D})$ ensures data confidentiality, integrity, and authenticity, while the PUF-based authentication operations in LA guarantee the authenticity of $(Z, t)$. For the on-chain verification and snapshot structure, \janus\ ensures that the channel attackers cannot tamper with the messages. Moreover, our UC-based security proof shows that the security strength of our protocols is bounded by PUF and cryptographic primitives.

$\bullet$\ \textit{DoS attacks}: As assumed in \cref{sec:threat_model}, the DoS attacks here are similar to those in other works \cite{petziSCRAPSScalableCollective2022,ankergardPERMANENTPubliclyVerifiable2022}, referring to $\adver$ sending massive meaningless and erroneous attestation requests to a legitimate $\att$ to exhaust the device resources. \janus\ can mitigate this attack in three ways: (1) The blockchain is naturally a robust distributed system that withstands numerous data requests. (2) If the system is overwhelmed, the duplex architecture of \janus\ allows users to opt for off-chain protocols, or the other way around. This resilience helps guarantee the continuous service delivery. (3) Meanwhile, in our protocol design, most operations are based on lightweight primitives, \eg, AEAD algorithm, thereby imposing no excessive computational burdens on $\att$.


$\bullet$\ \textit{Physical probing}: If $\adver$ physically acquires the attested devices, it can repeatedly trigger the RA procedure to generate side-channel power leakage to deduce the key, or it directly dumps the binary in external memory to find the secret information. However, in our design, $K_{\gid}$ is constructed through PUF-based XOR masking. It can remain secure under the semi-invasive attack manners even if $\mli{MK}_{\aid}$ is exposed by power side-channel analysis or memory read-out. Also, our PUF-based RA protocol \cref{sec:remote_attestation} avoids using a fixed encryption key on the attester side in each session, thereby preventing the highly correlated power leakage.

$\bullet$\ \textit{Chip delayering}: Furthermore, if $\adver$ adopts invasive physical attacks such as FIB and LVP to reveal the electrical layout inside the TEE chip, such destructive attack would unavoidably damage the original structure of the embedded PUF instances. Therefore, even if $\adver$ manages to extract the TEE root key, it will be unable to generate valid attestation reports and pass the verification without PUF. Most RA designs cannot resist this type of attacks since they rely on a centralized TEE root key instead of physics-based trust.

$\bullet$\ \textit{PUF modeling attacks}: The communication messages in RA protocols and smart contracts do not expose raw PUF CRPs during transmission, preventing $\adver$ from gathering them to train a PUF model. Moreover, PUF's I/O bus is usually sealed in the chip \cite{nxplpcpuf,xilinxpuf} and thus not exposed to the outside. As a result, $\adver$ cannot directly invoke PUF unless it uses invasive measures, which, as previously explained, would alter the original IC status and turn PUF unavailable. Also, \janus\ is PUF-agnostic and can use various PUF designs \cite{XuTransferResilientPUF,NassarANVPUF, LinZPZPUF} with better modeling resilience.

$\bullet$\ \textit{Rogue verifier and collusion}: The decentralized attestation contract in \cref{sec:attestation_contract} prevents rogue verifiers from intentionally declaring the false attestation results. For off-chain protocols, even if there are corrupted verifiers that produce untrusted results, our audit contract helps identify such compromises and minimize the impact that rogue verifiers bring.




\section{Implementation and Evaluation}\label{sec:evaluation}

\begin{table*}[t]
    \footnotesize
	\centering
	\caption{Performance of off-chain remote attestation of \janus\ \label{tab:offchain_concrete_performance}}
    \begin{threeparttable}
	\begin{tabular}{cccccccccc}
		\toprule
		\multirow{2}{*}{\textbf{Items}} & \multicolumn{3}{c}{\textbf{Local Attestation}} & \multicolumn{3}{c}{\textbf{Remote Attestation}} & \multirow{2}{*}{\makecell[c]{\textbf{Binary}\\\textbf{size (KB)}}} & \multirow{2}{*}{\makecell[c]{\textbf{PUF}\\((8,8)-IPUF)}} & \multirow{2}{*}{\makecell[c]{\textbf{Communication}\\ \textbf{Latency}}} \\
        \cmidrule(r){2-4}\cmidrule(r){5-7}
        ~ & Overall & Challenge & Response & Overall & Challenge & Response &  \\
		\midrule
        \textbf{SGX Server} & 1.18 ms & 0.12ms & 0.81 ms & 4.98 ms & 1.64 ms (\ding{174}) & \makecell[c]{0.04 ms (\ding{173}) \\ 2.16 ms (\ding{177})}  & 548 & \multirow{2}{*}{\makecell[c]{2267 LUT +\\295 FF, \\ 14ms$^\dagger$}} & \multirow{2}{*}[-1ex]{\makecell[c]{3 ms $\sim$\\ 7 ms}} \\
        \textbf{LPC55S69} & \multicolumn{3}{c}{-} & 510.46 ms & \makecell[c]{3.47 ms (\ding{172}) \\ 164.12 ms (\ding{175})} & 341.98 ms (\ding{176}) & 348 & ~ & ~ \\
		\bottomrule
	\end{tabular}
    \begin{tablenotes}
        \small 
        \item $\dagger$: including the latency of serial port communication;
    \end{tablenotes}
    \end{threeparttable}
\end{table*}

\subsection{\janus\ Implementation}\label{sec:implementation}

We provide a PoC implementation of \janus\ in this paper. For off-chain attestation protocols, we use a LPC55S69 evaluation board as the verifier and an Dell EMC R750 Server as the attester. LPC55S69 enables ARM TrustZone and has dual core Arm Cortex-M33 up to 150 MHz. It is a widely used general-purpose IoT device \cite{NXPLPC55S6x}. The EMC Server is with 512 GB RAM and with Intel Xeon Gold 6330 3.1 GHz processor enabling SGX. We set an (8,8)-IPUF \cite{nguyen2019interpose} instance on a Digilent Nexys board with Xilinx Artix-7 embedded. IPUF is connected to the EMC Server to be the second RoT. The LPC board communicates with the EMC server using socket API. They simulate the scenario where distributed application modules require mutual RA.

For on-chain smart contracts, we use Hyperledger Sawtooth \cite{olsonSawtoothIntroduction2018} to implement our smart contracts in Docker containers, as in \cite{petziSCRAPSScalableCollective2022, ankergardPERMANENTPubliclyVerifiable2022}, The contracts are deployed on the EMC server and thus will be running with the attester code. We allow remote devices to access the Sawtooth network via REST API and the exposed container ports. Then the LPC board can interact with the Sawtooth blockchain.

%


\subsection{Prototype Evaluation}\label{sec:performance_eval}


\subsubsection{Off-Chain Protocol Evaluation}
Table \ref{tab:offchain_theoretical_performance} compares the theoretical performance of the PUF-based RA protocol in \cref{sec:remote_attestation} with SGX and LIRA-V \cite{shepherdLIRAVLightweightRemote2022}.






\noindent\textbf{Theoretical comparison}: We can see that \janus\ has a compact protocol structure using three-round interactions to complete mutual attestation. Compared to other protocols, \janus\ only needs asymmetric primitives in measurement signing, resulting in lower computation overhead on resource-constrained platforms. And \janus\ is the only one to achieve UC-based provable security. Also, \janus\ does not necessitate the pre-storage of cryptographic materials for peer-to-peer communication, reducing storage overhead to constant complexity. Therefore, the off-chain protocol of \janus\ is lightweight and scalable for numerous devices.

\begin{table}[H]
    \footnotesize
	\centering
	\caption{Comparison of PUF-based RA protocols\label{tab:offchain_theoretical_performance}}
	\begin{tabular}{ccccc}
		\toprule
		\textbf{Properties} & \textbf{Sanctum RA} \cite{costan2016sanctum} & \cite{RomanPUFHash} & \cite{AkramMMPUF} & \textbf{\janus} \\
		\midrule
		\textbf{RoT Type} & Key & Key & PUF & PUF \\
		\textbf{No. of RoT} & 1 & 1 & 1 & 2 \\ 
        \textbf{Mutual RA} & \colourxmark & \colourxmark & \colourxmark & \colourcheck \\
        \textbf{Security proof} & \colourxmark & \colourxmark & \colourxmark & \colourcheck \\
        \textbf{Signature calls} & 2 & 2 & 1 & 2\\
        \textbf{Key agreement} & ECDH & \tbNA & \tbNA & PUF XOR \\
        \textbf{Protocol round} & 2 & 2 & 4 & 3 \\
	\bottomrule
	\end{tabular}
\end{table}

\noindent\textbf{Microbenchmark:} The performances of the protocols are given in Table \ref{tab:offchain_concrete_performance}. We here choose ASCON AEAD \cite{dobraunig2021ascon}, SHA256 and ECDSA with secp256r1. The devices genuinely execute the PUF-based RA protocol ($\sim$1000 LoC) meanwhile we use \texttt{gettimeofday()} and hardware cycle counter \texttt{KIN1\_DWT\_CYCCNT} to record the latency.

The local attestation protocol ($\sim$1500 LoC) involving no asymmetric cryptographic operations can be completed in less than 1.2 ms on SGX enclaves in the server. The majority of computation latency for the RA protocol comes from the signing and verifying operation that is about 157.7 ms and 337.30 ms on the LPC board. The AEAD-based design can save considerable computation resources. Also, the raw executable files are 548 KB and 348 KB on each platform, making it resource-friendly and highly portable.




\subsubsection{On-Chain Smart Contract Evaluation}

We implement the smart contracts in Python ($\sim$600 LoC) and the corresponding client in C ($\sim$800 LoC) that allows devices to form and send the transactions to Sawtooth Network.

\noindent\textbf{Microbenchmark:} Table \ref{tab:onchain_concrete_performance} summarizes the overall performance of on-chain attestation of \janus. We measure the performance from two perspectives: (1) client preparation and (2) contract execution. For (1), except the attestation data generation, the required operations by Sawtooth including data serialization and transaction wrapping bring much overhead actually. Hence, as shown in Table \ref{tab:onchain_concrete_performance}, the computation latency on the board is about 300 ms$\sim$400 ms and the server is about 3 ms$\sim$4 ms. For (2), off-chain data submissions are executed faster than transaction payload verification. The submission latency of three contracts are all less than 5 ms and is acceptable for real-time systems. Our tests also find out that the LPC board requires about 12 ms to retrieve the stored data on-chain through unix sockets in local network. These results show that our off/on-chain interactions do not adversely affect the overall performance of the entire RA procedure in \janus.

\subsubsection{Scalability Evaluation}

We present the scalability evaluation of \janus's off-chain and on-chain attestation


\noindent\textbf{Baselines:} We choose \cite{zhengPUFbasedMutualAuthentication2022}, \cite{shepherdLIRAVLightweightRemote2022} as the off-chain comparison baseline and \cite{petziSCRAPSScalableCollective2022} for on-chain. \cite{zhengPUFbasedMutualAuthentication2022} presents an PUF-based authentication protocol and \cite{petziSCRAPSScalableCollective2022} also uses smart contract for attestation and thus is partially similar with \janus's on-chain functions.

\begin{figure}[htbp]
\centering
\subfloat[Off-chain RA protocol]{%
  \includegraphics[clip,width=0.78\columnwidth]{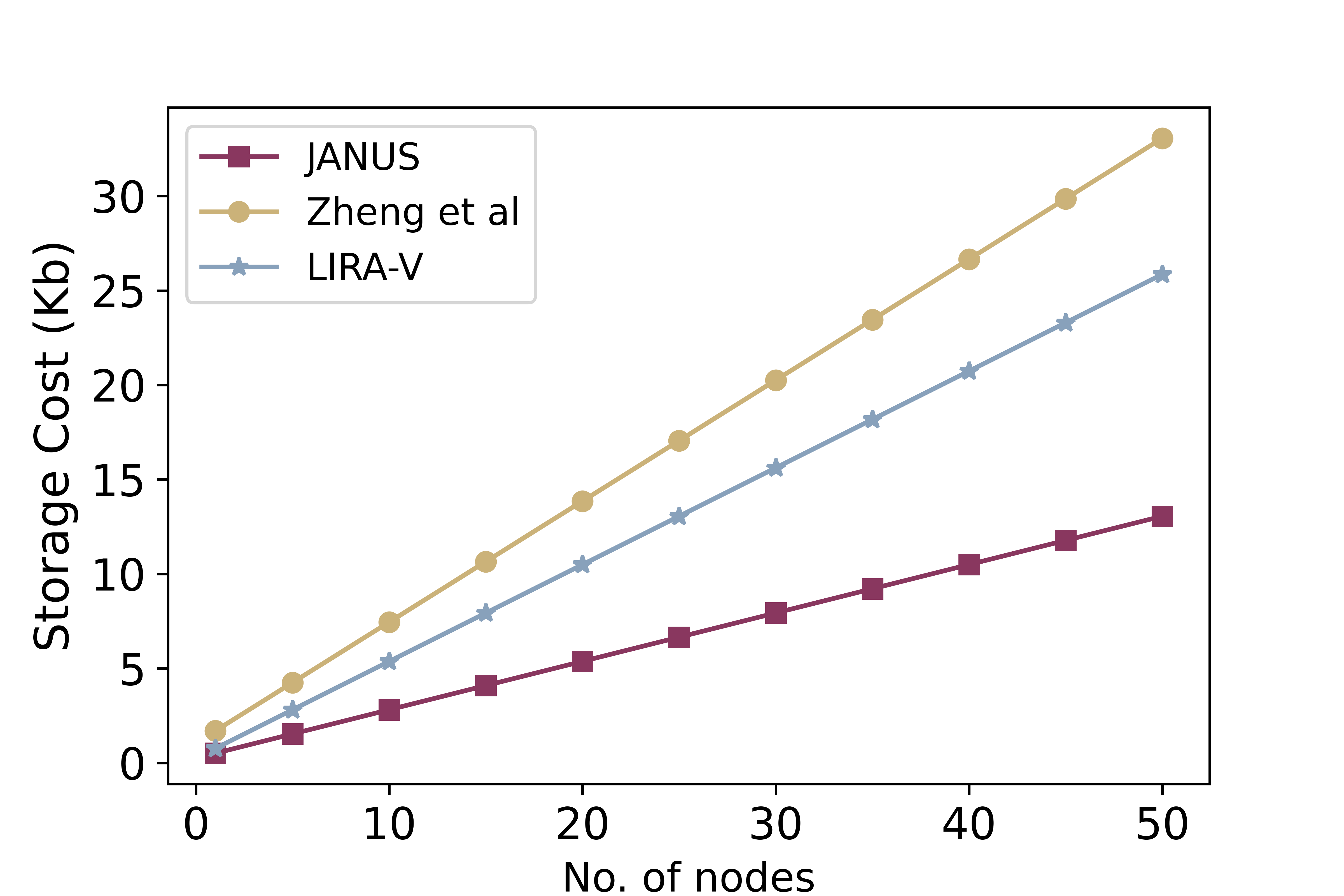}\label{fig:scalability_evaluation_offchain}%
}\\[-1.2ex]
\subfloat[On-chain Attestation]{%
  \includegraphics[clip,width=0.78\columnwidth]{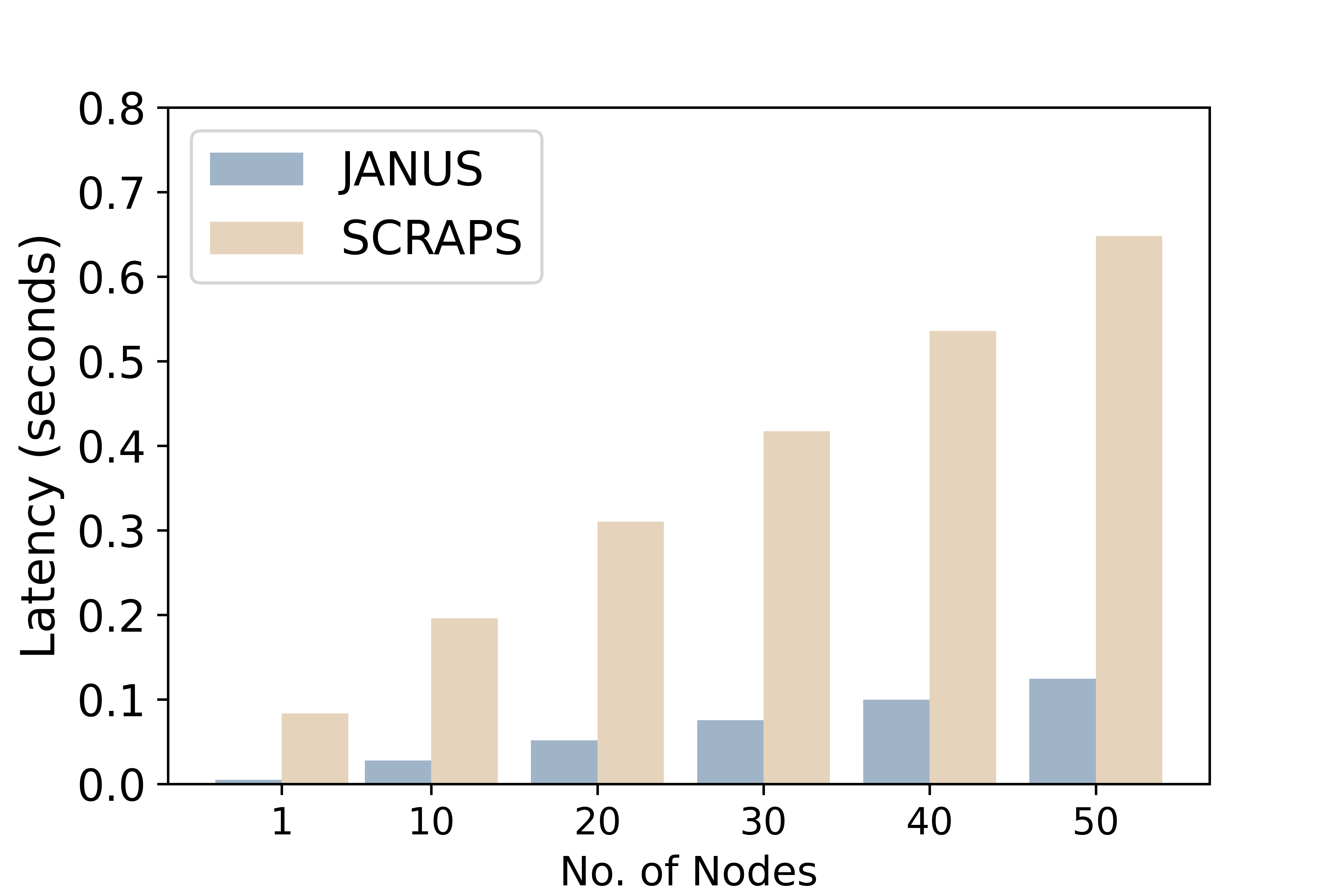}\label{fig:scalability_evaluation_onchain}%
}
\caption{Scalability evaluation of \janus\ when increasing the involved attested participants}
\label{fig:scalability_evaluation}
\end{figure}

\noindent\textbf{Comparison:} Figure \ref{fig:scalability_evaluation_offchain} illustrates the storage cost incurred by a $\vrf$ when communicating with other $\att$s. Compared with \cite{zhengPUFbasedMutualAuthentication2022, shepherdLIRAVLightweightRemote2022}, \janus\ requires the minimal cryptographic materials in E2E communication. Figure \ref{fig:scalability_evaluation_onchain} provides a comparison of on-chain attestation latency between \janus\ and SCRAPS \cite{petziSCRAPSScalableCollective2022}. \janus\ exhibits a slower upward trend as the number of $\att$ increases and save many interaction latency than SCRAPS. \janus\ can offer better scalable off-chain and on-chain attestation services for large-scale networks.

\begin{table*}[htpb]
	\centering
	\caption{Performance of onchain attestation of \janus\ \label{tab:onchain_concrete_performance}}
	\begin{tabular}{cc|cc|cc|cc}
			\toprule
			\textbf{Side} & \textbf{Item} & \multicolumn{2}{c|}{\textbf{Attestation}} & \multicolumn{2}{c|}{\textbf{Audit}} &  \multicolumn{2}{c}{\textbf{Attestation Switch}} \\
			\midrule
			\multirow{4}{*}{\textbf{Contract}} & Script size & \multicolumn{2}{c|}{8.0 KB} & \multicolumn{2}{c|}{8.0 KB} & \multicolumn{2}{c}{5.7 KB} \\
			~ & Off-chain Data submission & \multicolumn{2}{c|}{2.5 ms} & \multicolumn{2}{c|}{ 3.7 ms} & \multicolumn{2}{c}{4.4 ms} \\
			~ & TX Payload Verification & \multicolumn{2}{c|}{29.6 ms} & \multicolumn{2}{c|}{ 8.3 ms } & \multicolumn{2}{c}{-} \\
			~ & Total calls & \multicolumn{2}{c|}{3} & \multicolumn{2}{c|}{2} & \multicolumn{2}{c}{2} \\
			\hline
            \hline
			\multirow{3}{*}{\textbf{Client}} & \textbf{Platform} & Challenge & Attestation report & Credentials & Request & $\mathsf{dc}$ update & $\mathsf{as}$ submission \\
			~ & EMC SGX Server & 3.15 ms & 3.75 ms & 3.11 ms & 3.08 ms & 3.06 ms & 3.17 ms \\
			~ & LPC55S69 & 341.26 ms & 486.25 ms & 333.81 ms & 324.75 ms & 325.85 ms & 338.60 ms \\
			\bottomrule
		\end{tabular}
\end{table*}

\section{Discussion}\label{sec:discussion}

In this section, we further discuss why the purpose of choosing PUF in this work is different from others, and additional efforts of integrating PUF into real-world TEEs.


\subsection{Decentralized Trust of PUF for TEE}\label{sec:rationale_puf}


PUF internalizes environmental randomness on its structure. It provides a way to create physical intrinsic trust, which is desired in current TEEs. The following characteristics of PUF make it an essential component for TEE RA.

\begin{itemize}[itemsep=0pt, parsep=0pt, topsep=0pt]
    \item \textbf{New security guarantees}: TEE, TPM and HSM use the same way of ensuring the security, \ie, sealing the fused private keys. However, PUF introduces a new security paradigm by the unique intrinsic structures instead of a given key. It eliminates the target, \eg, keys, of many attacks and thus offers new and stronger guarantees in addition to the sealing key-based approach.
    \item \textbf{Independent trust}: PUF reserves an untouched area in circuits that cannot be modified by others (even its manufacturer) once the fabrication is completed. Therefore, it creates a new source of trust that is truly independent. Compared to other PUF-based works, we provide this insight to further exploit its capabilities, which is beyond than the PUF-backed key generation or physical hashing.
    \item \textbf{Succinct usage}: PUF only exposes challenge and response port. Compared with TPM and other techniques, PUF has a quite succinct usage which benefits the integration into many security designs. 
\end{itemize}

The succinct usage of PUFs leads to apparent similarities between some PUF-based RA designs \cite{AkramMMPUF,PReFeRPUFAtt,kongPUFattEmbeddedPlatform2014} and \janus. The fundamental difference lies in that we address the long-standing issue of trust establishment in TEE RA by leveraging the independent trust and new security guarantee of PUFs, which has not been well explored by other studies.












\subsection{Integrating PUF into TEEs}\label{sec:integrate_puftee}

Intel \cite{intelpuf}, NXP \cite{nxplpcpuf} and Xilinx \cite{xilinxpuf} have released products with integrated PUF. With PUF being standardized \cite{PUFISOP1,PUFISOP2}, a hardware ecosystem with built-in PUF can be expected. Here we discuss a possible roadmap of integrating it into TEE.


The hardware integration of PUF in TEE would require xPU (\eg, CPU, GPU and NPU) extensions and new xPU-PUF interface design. Potential efforts include: (1) On the PUF side, it should be defined as an attestable fixed function module. It should be setup with a clear secure model (\eg, side-channel resistant). (2) On the xPU side, new hardware extensions or drivers should be added to cooperate with the installed PUF. Then TEE/RA related instructions (e.g., measurement or signing) could optionally be enforced by PUF. (3) The (off-die) physical link between xPU and PUF should be encrypted. It may also need access control and memory management. We leave the off-the-shelf implementation for future work.





\subsection{Limitations}

We identify two limitations of \janus. First, PUF here cannot operate alone to obtain measurements in case of TEE failures. It must work in TEE to generate the physical attestation endorsements in a nested way. We intend to address this limitation in the future. Second, \janus\ does not provide formally-verified smart contract functions including the attestation switch. Therefore, their security strength is difficult to be evaluated. We consider leveraging TEE to enhance the security and privacy of the contract execution \cite{SGXonerated}.







\section{Conclusion and Future Work}\label{sec:conclusion}

To decentralize the trust establishment in traditional TEE RA designs, we propose using PUF as an additional intrinsic RoT to break the closed RA architecture dominated by TEE manufacturers. Our PUF-based attestation protocols are lightweight in both computation and storage. And to eliminate the centralization in single-party verification, we design several smart contracts to build a publicly verifiable and permanently reviewable procedure. Our switch mechanism combines the different attestation functionalities of \janus\ to form a resilient RA service. \janus\ thoroughly addresses the limitations of over-centralization and lack of resilience that exist in nearly all TEE RA schemes. It can be a comprehensive solution of increasing trust in TEE and shed some light on designing a more secure and trusted TEE RA paradigm. 

In the future, we plan to overcome the aforementioned drawbacks of \janus. We will design a dedicated PUF-based measuring mechanism for different platforms including TEE. It is expected to achieve lightweight runtime measuring by PUF. Building on this and \janus\ framework, we will explore how to use PUF to provide runtime measurement and design PUF-based unified heterogeneous TEE systems.




\section*{Acknowledgments}
The authors would like to thank the editors and anonymous reviewers for their work. This work is supported by National Science and Technology Major Project (No. 2022ZD0120304) and Shanghai Science and Technology Innovation Action Program (No. 22511101300).



\bibliographystyle{IEEEtran}
\bibliography{janus.bib}

\begin{IEEEbiography}[{\includegraphics[width=1in,height=1.25in,clip,keepaspectratio]{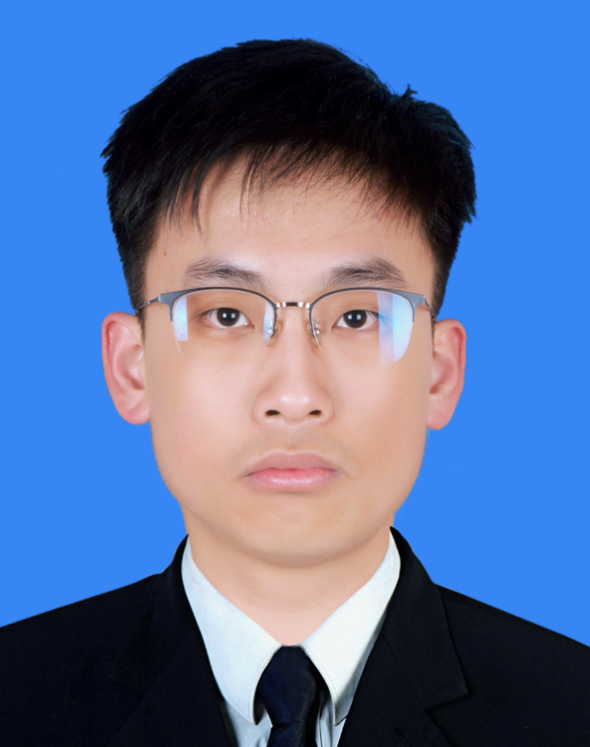}}]{Xiaolin Zhang} received from Xidian University of China, Xi'an, his B.S. degree in cyber security in 2020. He is currently a Ph.D. candidate under successive postgraduate and doctoral program at School of Electronic Information and Electrical Engineering, Shanghai Jiao Tong University, Shanghai. His research interests include design of PUF-based security schemes and cryptography. He has published scientific papers on IEEE TCAD.
\end{IEEEbiography}

\begin{IEEEbiography}[{\includegraphics[width=1in,height=1.25in,clip,keepaspectratio]{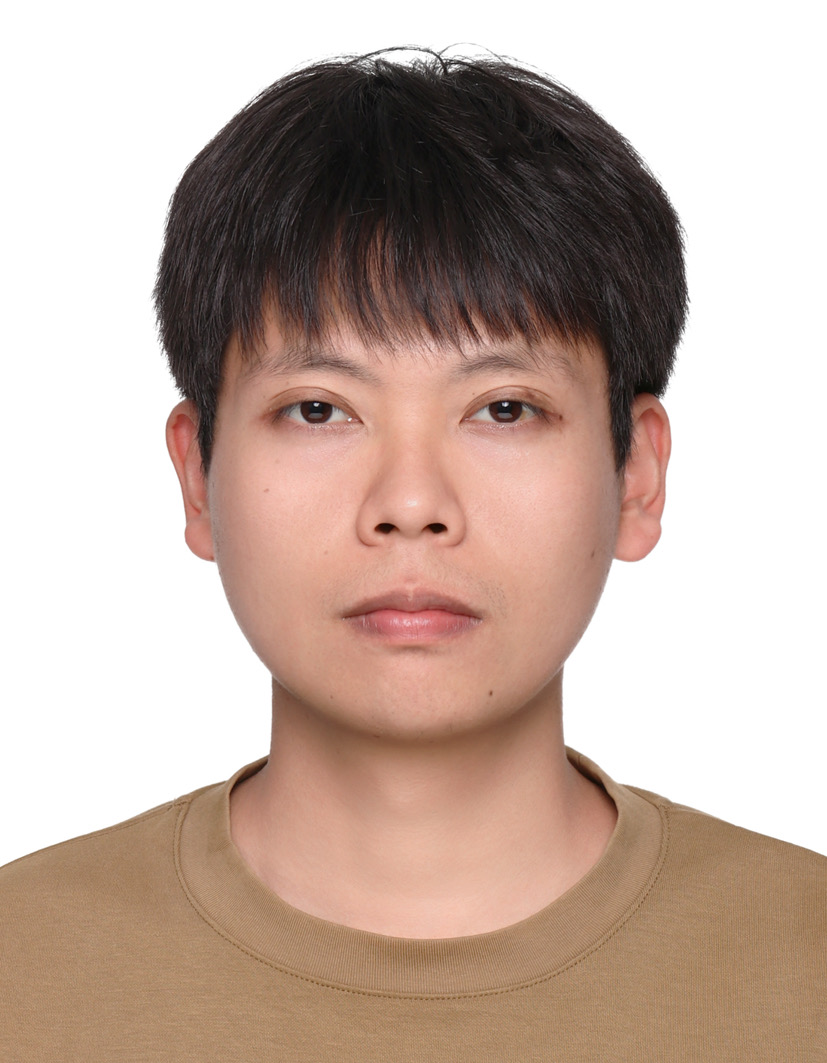}}]{Kailun Qin} is currently a DEng candidate at School of Electronic Information and Electrical Engineering, Shanghai Jiao Tong University (SJTU), Shanghai. He obtained his master's degrees respectively from IMT Atlantique (France) and Nanjing University of Science and Technology (China) in 2017. His research interests include confidential computing, trusted execution environment and system security. He has been an active speaker at multiple cloud and security industry conferences.
\end{IEEEbiography}

\begin{IEEEbiography}[{\includegraphics[width=1in,height=1.25in,clip,keepaspectratio]{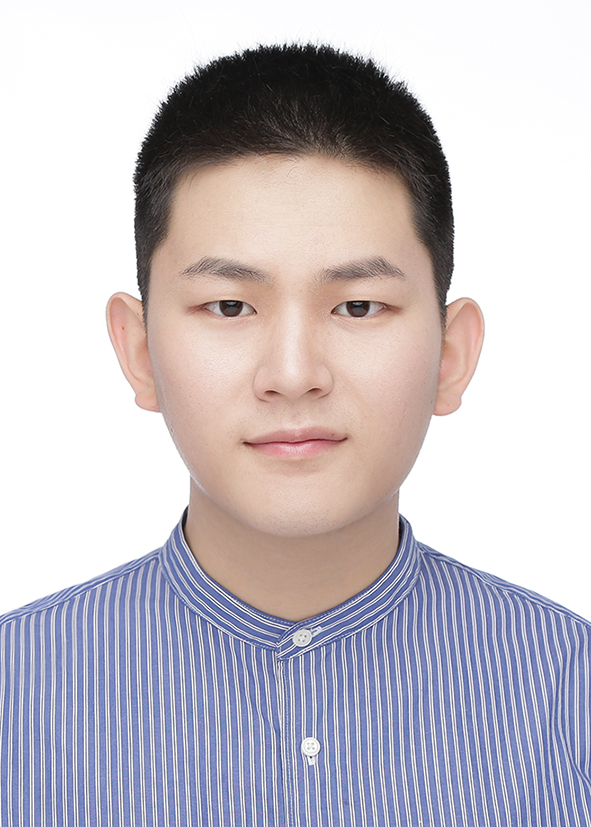}}]{Shipei Qu} received from Xidian University of China, Xi'an, his B.S. degree in Electronic Engineering in 2020. He is currently a Ph.D. candidate at School of Electronic Information and Electrical Engineering,Shanghai Jiao Tong University, Shanghai. His research interests include hardware security and cryptography.
\end{IEEEbiography}

\begin{IEEEbiography}[{\includegraphics[width=1in,height=1.25in,clip,keepaspectratio]{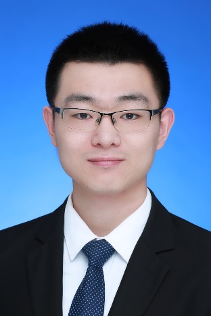}}]
  {Tengfei Wang} received the B.E. degree and the M.E. degree from Tianjin University, Tianjin, China, in 2016 and in 2019, respectively. He is currently pursuing the D.E degree from the School of Electronic Information and Electrical Engineering, Shanghai Jiao Tong University, Shanghai. His current research interests include circuit and system designs for cryptosystems. He has published scientific papers on IEEE TVLSI and other academic journals.
\end{IEEEbiography}

\begin{IEEEbiography}[{\includegraphics[width=1in,height=1.25in,clip,keepaspectratio]{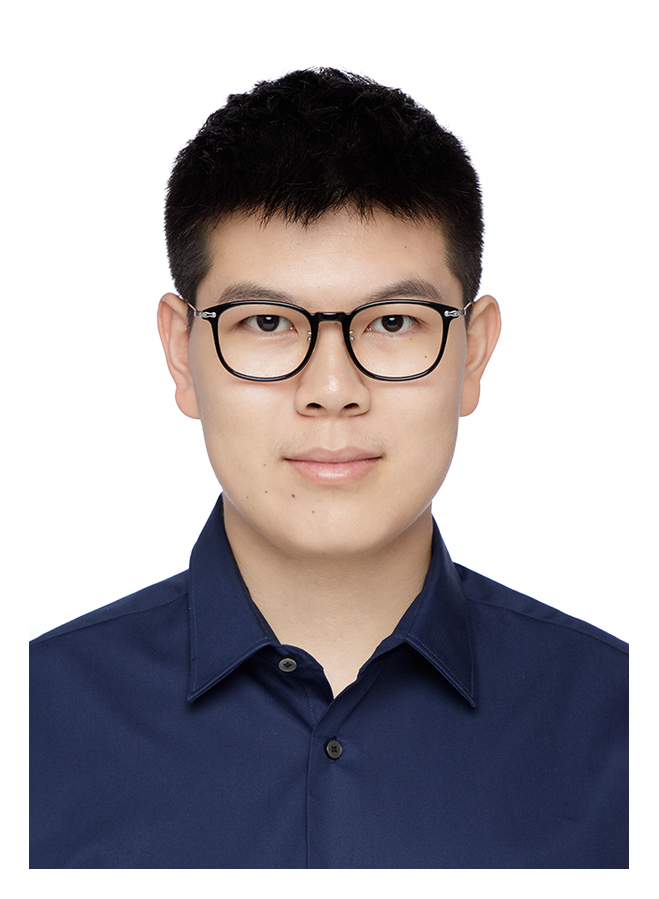}}]
  {Chi Zhang} is currently a research associate at School of Electronic Information and Electrical Engineering, Shanghai Jiao Tong University, Shanghai. He received the B.S. degree in computer science and technology from Southeast University in 2014, and the Ph.D. degree in computer science and technology from Shanghai Jiao Tong University in 2022. His research interests include cryptographic engineering, hardware security such as side-channel analysis and fault inject analysis, and cryptography. He has published scientific papers on TCHES, IEEE TVLSI, TIFS and DAC, AsianHost and other academic journals and conferences.
\end{IEEEbiography}

\begin{IEEEbiography}[{\includegraphics[width=1in,height=1.25in,clip,keepaspectratio]{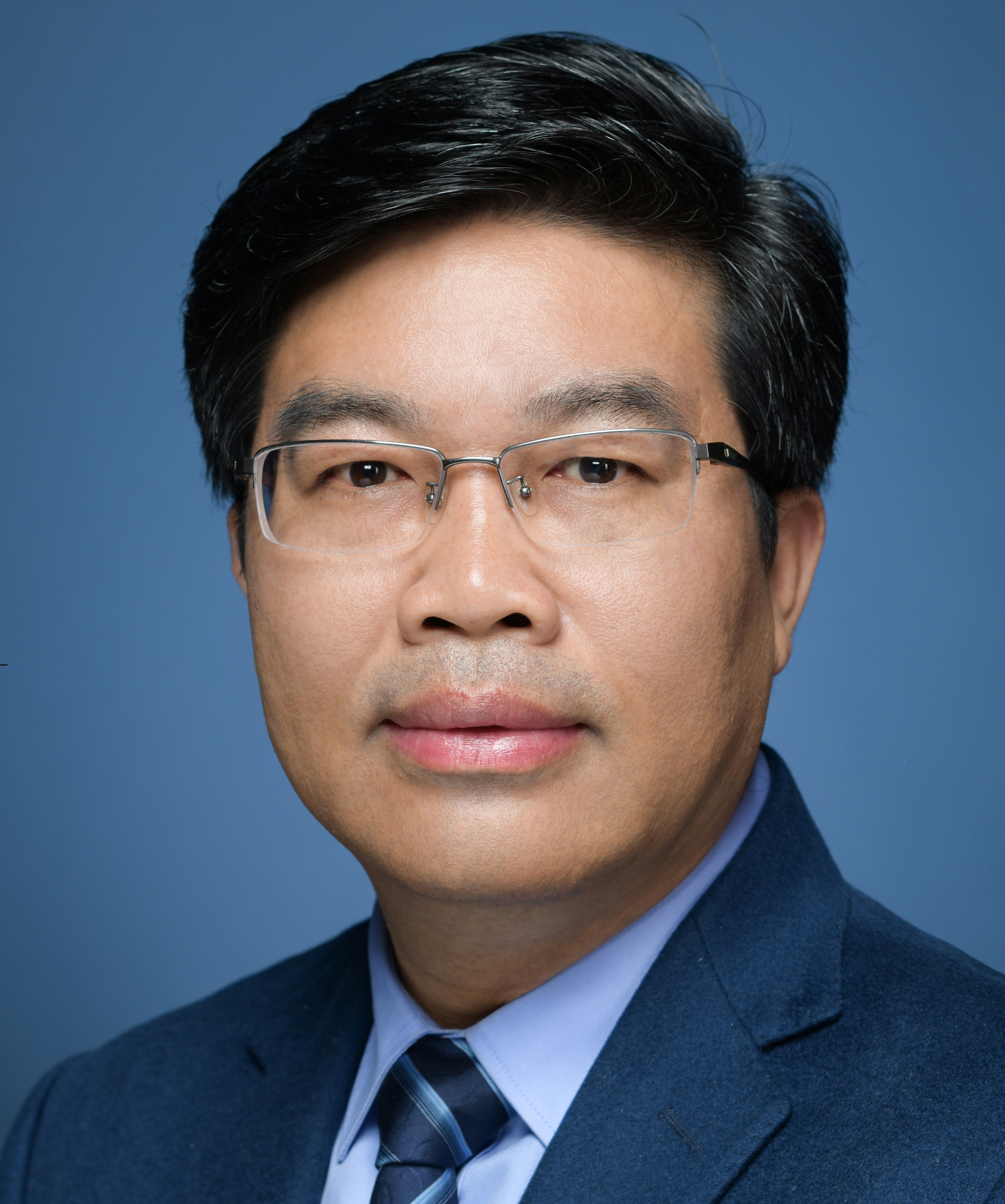}}]{Dawu Gu} is a distinguished professor at School of Electronic Information and Electrical Engineering, Shanghai Jiao Tong University (SJTU), Shanghai. He received from Xidian University of China, Xi'an, his B.S. degree in applied mathematics in 1992, and his M.S. degree in 1995 and Ph.D. degree in 1998 both in cryptography. His research interests include crypto algorithms, crypto engineering, and system security. He was the winner of Chang Jiang Scholars Distinguished Professors Program in 2014 by Ministry of Education of China. He won the National Award of Science and Technology Progress in 2017. He is the vice president of the Chinese Association for Cryptologic Research (CACR). He has got over 200 scientific papers in top academic journals and conferences, such as CRYPTO, IEEE S\&P, ACM CCS, NDSS, TCHES. He also served as PC members of international conferences such as ASIACRYPT, ASIACCS, ACNS, ISC, ISPEC, ICICS, Globecom, IEEE TrustCom, and NSS for more than 30 times.
\end{IEEEbiography}

\newpage

\appendices

\section{Security Proof}\label{sec:security_proof_full}

As mentioned in \cref{sec:security_proof}, the following definition formalizes the basic idea of UC-based security proof.

\begin{definition}(UC-security)
A protocol $\pi$ UC-realizes the ideal functionality $\mathcal{F}$, if for any PPT adversary $\Aa$ there exists a
simulator S such that no PPT environment machine $\mathcal{Z}$ exists to distinguish whether it contacts with $\pi$ and $\Aa$ in the real world or $\mathcal{S}$ and $\mathcal{F}$ in the ideal world. We have,
\begin{equation}
    REAL_{\pi, \Aa, \mathcal{Z}}\approx IDEAL_{\mathcal{F}, \mathcal{S}, \mathcal{Z}}
\end{equation}
where $\approx$ denotes the computational indistinguishability.
\end{definition} 

The formalized ideal functionality $\mathcal{F}_{\att}$ for mutual attestation protocols is shown in Figure \ref{fig:ideal_functionality_Fatt}. We omit the repeated entry check of tape records on $\aid, \vid, \ssid$ in $\mathcal{F}_{\att}$. Assuming the system is parameteraized with $1^{\lambda}$, we take the remote attestation protocol $\pi_{RA}$ as an example to prove Theorem \ref{theorem1} as follows.



\begin{figure*}
    \centering
    \begin{mybox}{Functionality: $\mathcal{F}_{att}$}
        \textbf{Initialization}:
        \begin{itemize}[nosep,leftmargin=*]
            \item Upon receiving $(\textsc{Measurement}, \id, \prog)$ from a party $P$, generate platform information $\Ii$ and obtain measurement $m:=\mathsf{memms}(\prog, \Ii)$, then let $\Ll[\id, \prog]:=(\mathsf{measured}, \Ii, m)$ and return $(\textsc{Measured}, \id)$ to $P$.
            \item Upon receiving $(\textsc{Provision}, \id)$ from $P$, 
            \begin{enumerate}[nosep, label=(\alph*)]
                \item generate the platform key set $\Kk$ and helper data $\Ww$, update $\Ll[\id, :].\mathsf{append}(\Kk, \Ww)$;
                \item let $\Ll[\id, :].\mathsf{state}=\mathsf{provisioned}$ and return $(\textsc{Provisioned}, \Ww, \id)$ to $P$. 
            \end{enumerate}
        \end{itemize}
        \textbf{Mutual Report}:
        \begin{itemize}[nosep,leftmargin=*]
            \item Upon receiving $(\textsc{Challenge}, i, j, \prog_{i}, \ssid)$ from $P_{i}$,
            \begin{enumerate}[nosep, label=(\alph*)]
                \item send the message to $\Ss$ and update $\Ll[i, \prog_{i}].\mathsf{append}(\ssid)$;
                \item if receive $(\textsc{Challenged}, A, C, T, \ssid)$ from $\Ss$, send it to $P_{j}$ and set $\Ll[j, \prog_{j}].\mathsf{state} = \mathsf{challenged}$.
            \end{enumerate}
            \item Upon receiving $(\textsc{MutReport}, i, j, \mathsf{prog}_{j}, \ssid)$ from $P_{j}$,
            \begin{enumerate}[nosep, label=(\alph*)]
                \item send the message along with $(\iid, \prog_{i}, \mathcal{I}_{j}, \prog_{j})$ to $\Ss$ and update $\Ll[j, \prog_{j}].\mathsf{append}(\ssid)$;
                \item if receive $(\textsc{MutAttReport}, \langle A_{i}, C_{i}, T_{i}\rangle, \langle A_{j}, C_{j}, T_{j}\rangle, \sigma_{i}, \sigma_{j}, \ssid)$ from $\Ss$, set $\Ll[i, \prog_{i}].\mathsf{state} = \mathsf{challenged}$. Then send $(\textsc{Attestation}, A_{i}, C_{i}, T_{i}, \mathsf{ssid})$ and $(\textsc{Attestation}, A_{j}, C_{j}, T_{j}, \mathsf{ssid})$ to $P_{i}, P_{j}$, respectively.
            \end{enumerate}
        \end{itemize}
        \textbf{Mutual Verification}:
        \begin{itemize}[nosep,leftmargin=*]
            \item Upon receiving $(\textsc{Verification}, i, j, A_{i}, C_{i}, T_{i}, A_{j}, C_{j}, T_{j}, \mathsf{ssid})$ from $P_{i}$ or $P_{j}$, 
            \begin{enumerate}[nosep, label=(\alph*)]
                \item obtain the signature $\sigma_{i}:=\mathsf{Recov}_{\kid, \wid}(A_{i}, C_{i}, T_{i})$ and $\sigma_{j}:=\mathsf{Recov}_{\Kk_{j}, \Ww_{j}}(A_{j}, C_{j}, T_{j})$; 
                \item if the party is corrupted, or the submitted record and the signature have not appeared or appear more than once, set $f[i, j]:=1$ or $f[j, i]:=1$ accordingly; otherwise let $f[i, j]$ or $f[j, i]$ be the result of $\mathsf{PrfVrf}_{\Kk, \Ww}(\sigma, \Ii, m)$;
                \item set $\Ll[i, \prog_{i}].\mathsf{state}=\Ll[j, \prog_{j}].\mathsf{state}=\mathsf{attested}$ and send $(\textsc{Attested}, f[i, j], \mathsf{ssid})$, $(\textsc{Attested}, f[j, i], \mathsf{ssid})$ to $P_{i}, P_{j}$, respectively.
            \end{enumerate}
        \end{itemize}    
    \end{mybox}
    \vspace{9pt}
    \caption{Ideal Functionality of Mutual Attestation}
    \label{fig:ideal_functionality_Fatt}
\end{figure*}

\begin{proof}
    We construct a simulator $\Ss$ that precisely emulates $\Aa$'s behaviors. We assume that a PPT adversary $\Aa$ can statically corrupt any party in $\pi_{RA}$, \ie, $\vrf$ or $\att$, to obtain its internal state and tape input and output. $\Ss$ runs an internal copy of $\Aa$ and then proceeds in the following cases.
    
\noindent \textbf{Case 1.} \textit{$\att$ and $\vrf$ are both honest.}

$\Aa$ in the real world only obtains the channel messages if $\att$ and $\vrf$ are not corrupted. So $\Ss$ in the ideal world need to simulate the messages exchanged between $\att$ and $\vrf$ in the real world. Upon receiving \textsc{Provision} from $\mathcal{F}_{\att}$, $\Ss$ randomly selects $\Kk_{\att} := (MK_{\id}, K)$ and $\Kk_{\vrf} := (s_{\id}, S)$. Upon receiving \textsc{Challenge}, $\Ss$ obtains $R_{V}, R_{A}, R_{T}\sample\bin^{n}$ and simulates \textsc{MutAttReport} by randomly sampling PUF-based measurement $RM$ and attestation proof $\sigma$. Then it uses AEAD encryption $\mathcal{E}$ to get $(A,C,T)$ for $\att$ and $\vrf$. It sends these messages to the internal adversary $\Aa$ who outputs them to the environment machine $\mathcal{Z}$. We denote $\mathsf{Adv}_{\pi_{RA}, \mathcal{F}_{att}}^{\mathrm{IND}}(Z)$ as the probability upper bound of $Z$ distinguishing them with the real-world messages, and we have $\mathsf{Adv}_{\pi_{RA}, \mathcal{F}_{att}}^{\mathrm{IND}}(Z) \le 3\mathsf{Adv}_{\mathrm{AEAD}}^{\mathrm{IND}}(Z) + \mathsf{Adv}_{\mathrm{PUF}}^{\mathrm{IND}}(Z) + 2\mathsf{Adv}_{\mathrm{Sig}}^{\mathrm{EUF}}(Z) + 2\mathsf{Adv}_{\mathrm{H}}^{\mathrm{Coll}}(Z) = \epsilon(\lambda)$ where $\epsilon(\lambda)$ is a negligible value ($negl$) related to $\lambda$.

\noindent \textbf{Case 2.} \textit{ $\att$ is corrupted and $\vrf$ is honest.}

If $\Aa$ corrupts $\att$ in the real world, $\Ss$ can also obtain the historical states of $\att$. Therefore, $\Ss$ can get $\Kk_{\att}$ to decrypt the old $(C, T)$ to obtain the measurement $RM$ and recover the attestation proof $\sigma$ during the challenge phase. Then, it can use these extracted materials to calculate $\att$'s output as if they are generated by $\Aa$ in the real world execution. However, since $\Ss$ cannot obtain the states of $\vrf$, it has to simulate $\vrf$'s output to $\mathcal{F}_{att}$, as in the \textbf{Case 1}. Consequently, the advantage of $\mathcal{Z}$ to distinguish $\vrf$'s output should not exceed $2\mathsf{Adv}_{\mathrm{AEAD}}^{\mathrm{IND}}(Z) + \mathsf{Adv}_{\mathrm{Sig}}^{\mathrm{EUF}}(Z) + \mathsf{Adv}_{\mathrm{H}}^{\mathrm{wColl}}(Z)= \epsilon(\lambda)$.

\noindent \textbf{Case 3.} \textit{$\att$ is honest and $\vrf$ is corrupted.}

Similar to \textbf{Case 2}, if $\Aa$ corrupts $\vrf$ in the real world, $\Ss$ can obtain $(s, S)$ to recover its measurement $M$ and the corresponding proof. Then $\Ss$ simulates $\att$'s output and forms $(A, C, T)$ as the input to $\mathcal{F}_{att}$. Since $\att$ is assumed to be equipped with PUF, $\Ss$ has to randomly sample a PUF response as the measurement. $\mathcal{Z}$ needs to break the indistinguishability game of ideal PUF, thereby having the advantage $\mathsf{Adv}_{\pi_{RA}, \mathcal{F}_{att}}^{\mathrm{IND}}(Z) \le \mathsf{Adv}_{\mathrm{AEAD}}^{\mathrm{IND}}(Z) + \mathsf{Adv}_{\mathrm{Sig}}^{\mathrm{EUF}}(Z) + \mathsf{Adv}_{\mathrm{H}}^{\mathrm{wColl}}(Z) + \mathsf{Adv}_{\mathrm{PUF}}^{\mathrm{IND}}(Z)=\epsilon(\lambda)$.

\noindent \textbf{Case 4.} \textit{ $\att$ and $\vrf$ are both corrupted.}

This case is straightforward since $\Ss$ can extract all states in $\att$ and $\vrf$.


\end{proof}

Note that the local attestation protocol can also be proven under the same ideal functionality. Therefore, the full off-chain attestation procedure of \janus\ can achieve UC-security by directly applying the UC-composable theorem in \cite{canetti2001universally}.


\section{Compatibility Real-world TEE Remote Attestation}\label{sec:attestation_report_table}

\noindent\textbf{Compatibility}: \janus\ is designed to have a regular attestation procedure as we expected. It can be easily integrated with real-world RA schemes by changing the payload field. Table \ref{tab:realworld_tee_ra_summary} summarizes the attestation report formats of mainstream TEE RA schemes. Moreover, since the off-chain remote attestation protocol in \cref{sec:remote_attestation} preserves a symmetric structure for both parties, our protocol can be scalable and lightweight regardless of whether attesters and verifiers have PUF installed. For on-chain attestation, the smart contracts are loosely coupled with device hardware configurations. \janus\ focuses more on the architectural design of attestation and is open to underlying TEE specifications, making it compatible with different systems with heterogeneous devices.

Table \ref{tab:realworld_tee_ra_summary} summarizes the format of several real-world TEE RA designs, including Intel SGX, Azure and ACK-TEE (\textbf{A}libaba Cloud \textbf{C}ontainer Service for \textbf{K}ubernetes). This table can serve as the evaluation metric of the trust rating in \cref{sec:audit_contract}. Auditors can label each device based on the device configuration details in the table to reflect its trust confidence level. Additionally, this table can also serve as a reference for integrating \janus's attestation report into real-world TEE products.

\begin{table*}[t]
	\centering
	\caption{Real-world Remote Attestation Report Fields\label{tab:realworld_tee_ra_summary}}
     \renewcommand{\arraystretch}{2}
	\begin{tabular}{llllll}
		\toprule
        \textbf{Field} & \multicolumn{1}{c}{\textbf{Intel SGX}} & \multicolumn{1}{c}{\textbf{ACK-TEE}} & \multicolumn{1}{c}{\textbf{Azure for SGX}} & \multicolumn{1}{c}{\textbf{Azure for SEV-SNP}} & \multicolumn{1}{c}{\textbf{Azure for TPM}} \\
        \midrule
        \textbf{SVN} & \textsf{isvsvn, cpusvn} & \textsf{svn} & \textsf{svn} & \makecell[l]{\textsf{bootloader_svn, tee_svn}\\\textsf{guest_svn, microcode_svn}} & \textsf{tpmversion}\\
        \hline
        \textbf{Measurement} & \textsf{mrenclave} & \textsf{mrenclave} & \textsf{mrenclave} & \textsf{launchmeasurement} & - \\
        \hline
        \textbf{Signer} & \textsf{mrsigner} & \textsf{mrsigner} & \textsf{mrsigner} &  \textsf{authorkeydigest} & \textsf{aikPubHash} \\
        \hline
        \textbf{Signature} & \textsf{signature} & \textsf{signature} & \textsf{signature} & \textsf{signature} & \textsf{signature}\\
        \hline
        \textbf{Others} & \makecell[l]{\textsf{base_name},\\ \textsf{report_data, isvpoid}} & \textsf{product_id} & \makecell[l]{\textsf{product_id},\\ \textsf{report_data}} & \makecell[l]{\textsf{imageid, report_id}\\\textsf{report_data}} & \makecell[l]{\textsf{vbsReportPresent},\\\textsf{secureBootEnabled}} \\
		\bottomrule
	\end{tabular}
\end{table*}

\end{document}